\newif\ifpreprint
\preprinttrue 

\newif\ifSupplementary
\Supplementaryfalse
\ifpreprint
\documentclass{natureprintstyle}
\else
\documentclass{nature}
\fi
\usepackage{soul}
\usepackage{subfigure}
\usepackage{graphicx}
\usepackage{multirow}
\usepackage{amssymb,hyperref}
\usepackage{amsmath}
\usepackage{dblfloatfix}    
\usepackage{kantlipsum}     

\newcommand{\aj}{AJ}%
\newcommand{\apj}{ApJ}%
\newcommand{\apjl}{ApJL}%
\newcommand{\apjs}{ApJS}%
\newcommand{\aap}{A\&A}%
\newcommand{\mnras}{MNRAS}%
\newcommand{\pasp}{PASP}%
\newcommand{\nat}{Nature}%

\usepackage{multibib}
\newcites{maintext}{\mbox{ }}
\newcites{method}{\mbox{ }}
\newcites{supplementary}{\mbox{ }}

\DeclareMathOperator{\E}{\mathbb{E}}

\title{An intermediate-mass black hole in the centre of the globular cluster 47 Tucanae}

\author{B\"ulent K{\i}z{\i}ltan$^{1}$, Holger Baumgardt$^{2}$ \& Abraham Loeb$^{1}$}

\begin{document}

\maketitle

\begin{affiliations}
\item Harvard-Smithsonian Center for Astrophysics, 60 Garden St., Cambridge, MA 02138 USA
\item School of Mathematics and Physics, University of Queensland, Brisbane, QLD 4072,  Australia
\end{affiliations}

\begin{abstract}

Intermediate mass black holes play a critical role in understanding the evolutionary connection between stellar mass and super-massive black holes\cite{Volonteri:12}. However, to date the existence of these species of black holes remains ambiguous and their formation process is therefore unknown\cite{Baumgardt:05}. It has been long suspected that black holes with masses $10^{2}-10^{4}M_{\odot}$ should form and reside in dense stellar systems\cite{Sigurdsson:93,Ebisuzaki:01,Miller:02,Maccarone:07}. Therefore, dedicated observational campaigns have targeted globular cluster for many decades searching for signatures of these elusive objects. All candidates found in these targeted searches appear radio dim and do not have the X-ray to radio flux ratio predicted by the fundamental plane for accreting black holes\cite{Strader:12}. Based on the lack of an electromagnetic counterpart upper limits of $2060 M_{\odot}$ and $470 M_{\odot}$ have been placed on the mass of a putative black hole in 47 Tucanae (NGC 104) from radio and X-ray observations respectively\cite{de-Rijcke:06,Grindlay:01}. Here we show there is evidence for a central black hole in 47 Tuc with a mass of M$_{\bullet}\sim2300 M_{\odot}$$_{-850}^{+1500}$ when the dynamical state of the globular cluster is probed with pulsars. The existence of an  intermediate mass black hole in the centre of one of the densest clusters with no detectable electromagnetic counterpart suggests that the black hole is not accreting at a sufficient rate and therefore contrary to expectations is gas starved. This intermediate mass black hole might be a member of electromagnetically invisible population of black holes that are the elusive seeds leading to the formation of supermassive black holes in galaxies.

\end{abstract}

\begin{figure}[!h]
\centering
\includegraphics[trim =  .0mm .0mm 10.0mm 5.0mm, clip,width = 3.1in]{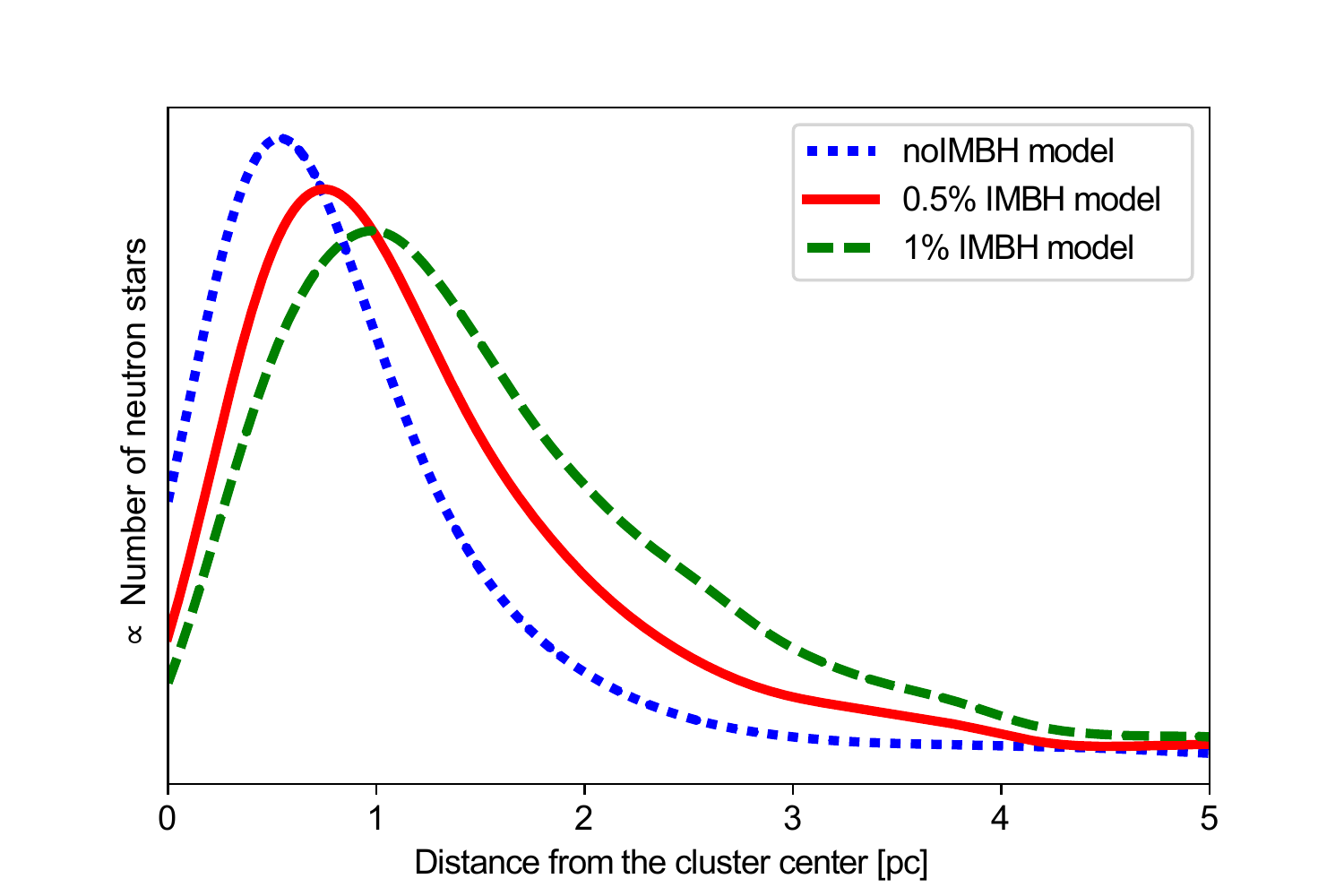}\\
\caption{{\bf Projected distribution of neutron stars in 47 Tuc.} Mass segregation is quenched by the black hole (IMBH models) and the density cusp is flattened. Due to this quenching, the number density of neutron stars in the inner most 1 pc is smaller when a black hole is present.}
\label{fig:segregation}
\end{figure}

An intermediate mass black hole (IMBH) strongly affects the spatial distribution of stars in globular clusters (GCs). Massive stars sink into the centre more efficiently during relaxation in order to achieve energy equipartition. As a consequence, as stars sink closer to the centre they are scattered by the black hole which heats up the core. This process quenches mass segregation (Figure \ref{fig:segregation}). Over the lifetime of a cluster, dynamical processes such as energy equipartition and two-body relaxation therefore contribute to the outward propagation of an integrated dynamical effect beyond the black hole's radius of direct influence. We find that this dynamical signature is efficiently propagated outward into the cluster. In relation to this effect, the distributions of pulsar accelerations for any projected distance from the centre show distinct features also sensitive to an IMBH. There are currently 25 millisecond pulsars detected within the cluster 47 Tuc. 19 of these have phase resolved timing solutions\cite{Freire:03,Ridolfi:16,Pan:16}(See Extended Data Table 1) which we use to infer spatial accelerations caused by the gravitational potential of the cluster\cite{Phinney:92}. Pulsar acceleration measurements together with $N$-body simulations provide stringent constraints on the mass of the central black hole in one of the most massive clusters (M$_{GC}\sim$$0.7\times10^{6}M_{\odot}$) with a compact core\cite{Baumgardt:16}.

\begin{figure}
\centering
\includegraphics[width = 3.in]{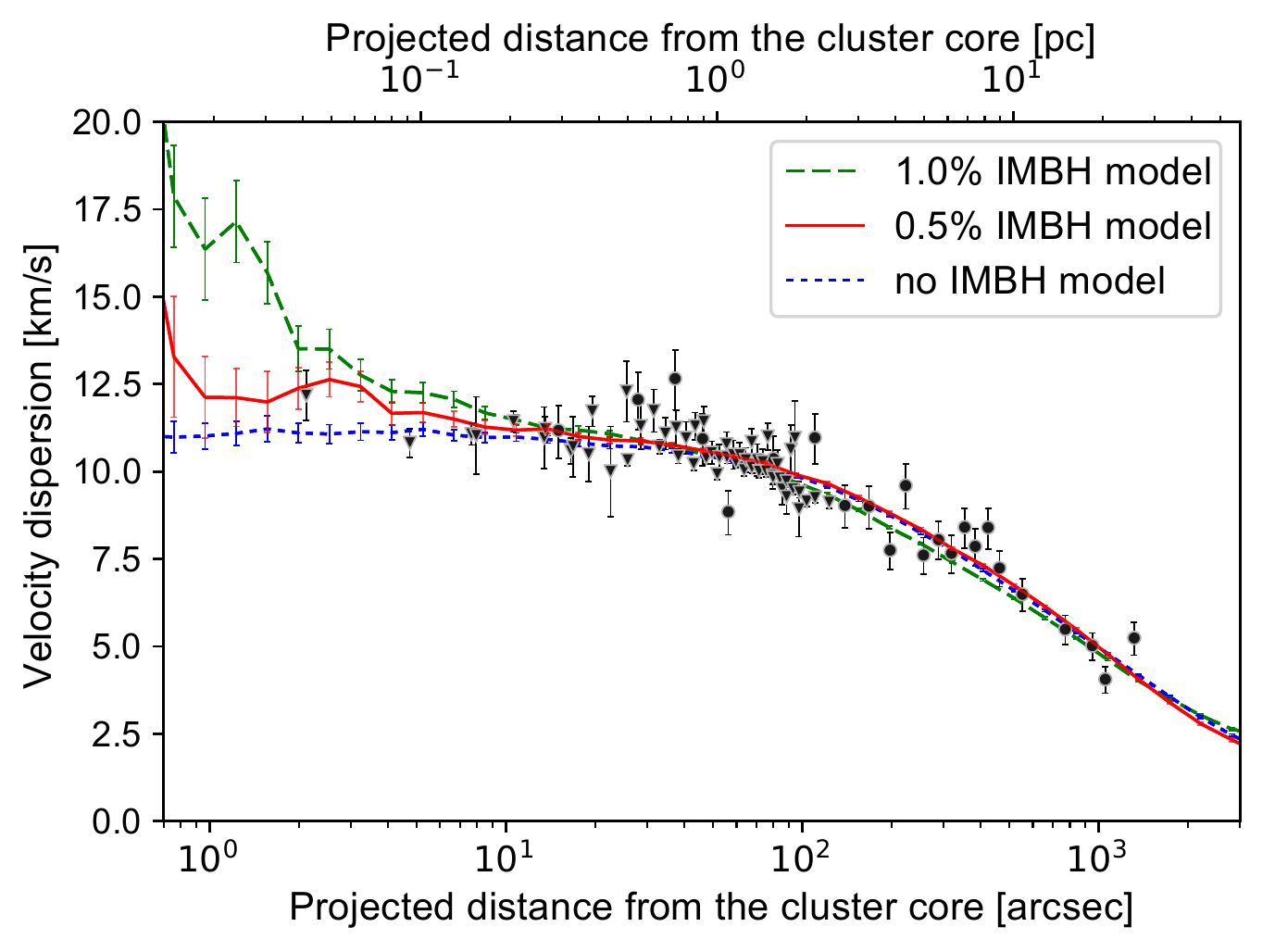}\\
\caption{{\bf Kinematic data for 47 Tuc compared with theoretical models.} The error bars are the 1-$\sigma$ dispersions from the Gaussian error distribution for each velocity measurement. Circle and triangle data points respectively denote radial velocity and proper motion observations\cite{Watkins:15}. The kinematic distance to 47 Tuc is d=$4$ kpc\cite{Baumgardt:16, McLaughlin:06}. We repeat the same calculations with models scaled to d$\sim4.5$ kpc (see footnote).}
\label{fig:velocity}
\end{figure}

IMBHs have distinct imprints on how massive stars dynamically settle three-dimensionally after the cluster has relaxed. The upper limit of M$_{\bullet}<1500 M_{\odot}$ placed on the black hole mass in 47 Tuc in earlier kinematic studies\cite{McLaughlin:06} appears inconclusive because new kinematic data\cite{Watkins:15} and $N$-body models\cite{Baumgardt:16} imply that no clear distinction between globular cluster models can be made based on available velocity dispersion measurements alone (Figure~\ref{fig:velocity}). In order to constrain the dynamical effects of a black hole in a cluster, we take a fundamentally different approach. In addition to pulsar accelerations, we jointly use this spatial imprint that carries information about the black hole beyond the radius of influence. We quantify how likely it is that a range of observed pulsar accelerations are related to specific model distributions. 

The dynamical $N$-body simulations of isolated star clusters evolve under the influence of stellar evolution and two-body relaxation. A grid of several hundred star clusters starting with different initial half-mass radii, density profiles and masses of their central black hole are run up to an age of T=11.75 Gyr to match the age of 47 Tuc\cite{Baumgardt:16}. We select those clusters that best match the surface density and velocity dispersion profile of 47 Tuc. The presence of primordial binaries does not play a significant role in the final segregation profile of heavy stars if clusters with the same surface density profile are compared (See Extended Data Figure 1 and Methods). No other a priori assumptions are made limiting the dynamics of the cluster.

The best-fitting models for the no-IMBH case and for IMBHs with 0.5\% and 1\% of the cluster mass are selected as a subset of viable replicas of 47 Tuc. Each $N$-body model considers full stellar evolution producing white dwarfs, neutron stars, and stellar mass black holes which are dynamically evolved along with the rest of the stars in the cluster. Models with black hole masses larger than 1\% of the cluster mass lead to fits which significantly deviate from the observed density and velocity dispersion profile of  47 Tuc. Hence such massive black holes are ruled out while models with smaller black hole masses or without a black hole lead to comparable fits.

In each simulated replica of 47 Tuc we have full spatial and kinematic information of all stars, including neutron stars. The majority of neutron stars ($\sim 90\%$) become unbound due to high kick velocities following the supernovae explosions during formation\cite{Pfahl:02,Ivanova:08}. We extract accelerations of all simulated neutron stars with projected distances that match the observed pulsars in 47 Tuc. Integrated accelerations of these matched neutron stars form distinctly different distributions (See Extended Data Figure 2).

\begin{figure}
\centering
\includegraphics[trim =  .0mm .0mm 0.0mm .0mm, clip,width = 3.in]{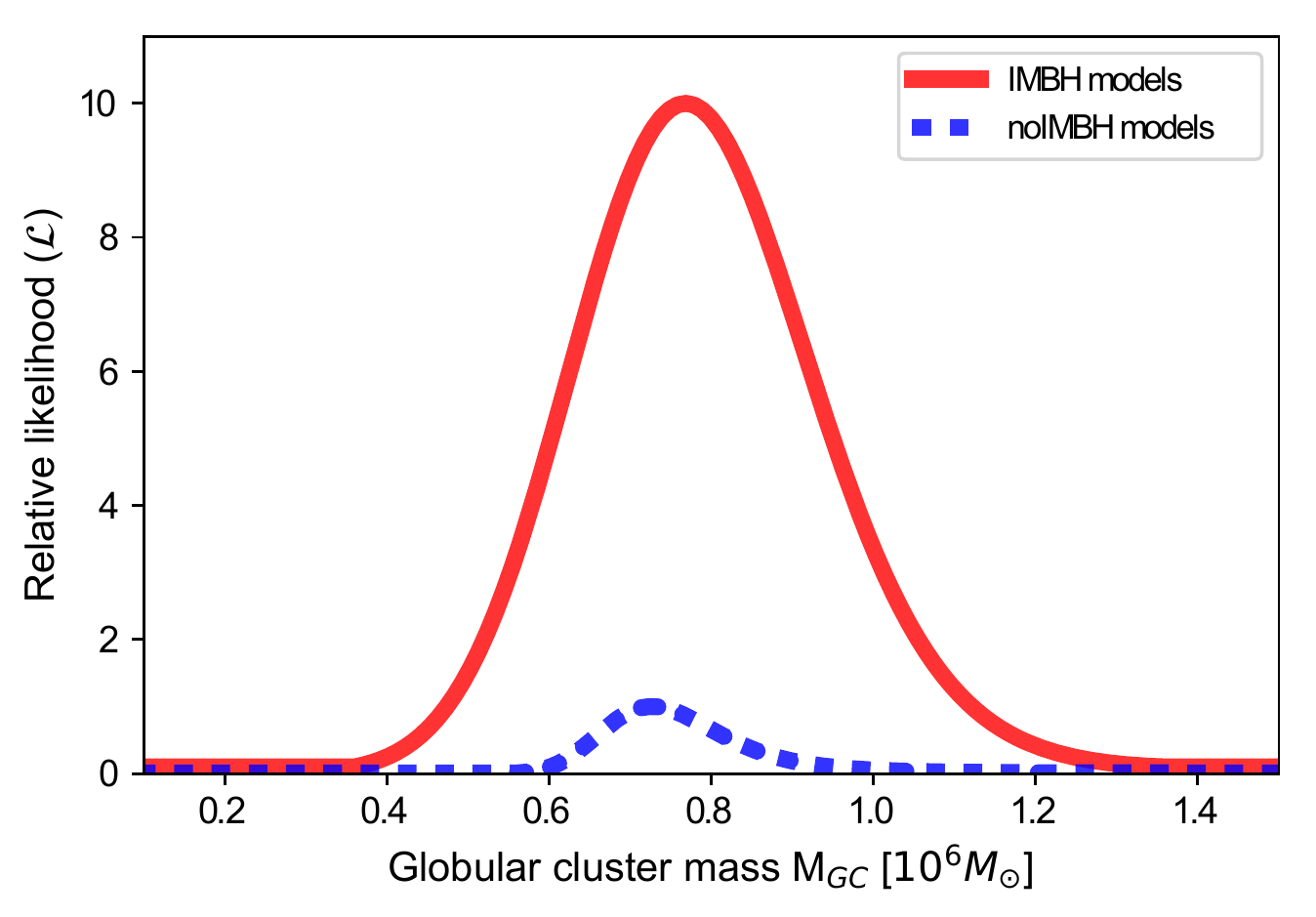}
\caption{{\bf Comparison of N-body model likelihoods of 47 Tuc.} The likelihood ($\mathcal{L}$) for each model is calculated by comparing the observed pulsar accelerations ($\mathcal{P}$) with the acceleration distributions produced in N-body simulations ($\mathcal{N}$) with and without an IMBH (see Equation~\ref{Eq:KL}).
The peak likelihood gives an independent dynamical measure of the total cluster mass M$_{GC}\sim0.76\times10^{6} M_{\odot}$. The observed pulsar accelerations and distributions within the cluster are at least 10 times more likely to be produced in the presence of a black hole.}
\label{Fig:models}
\end{figure}

\begin{figure*}[!t]
  \centering
  \subfigure[]{\includegraphics[scale=0.57]{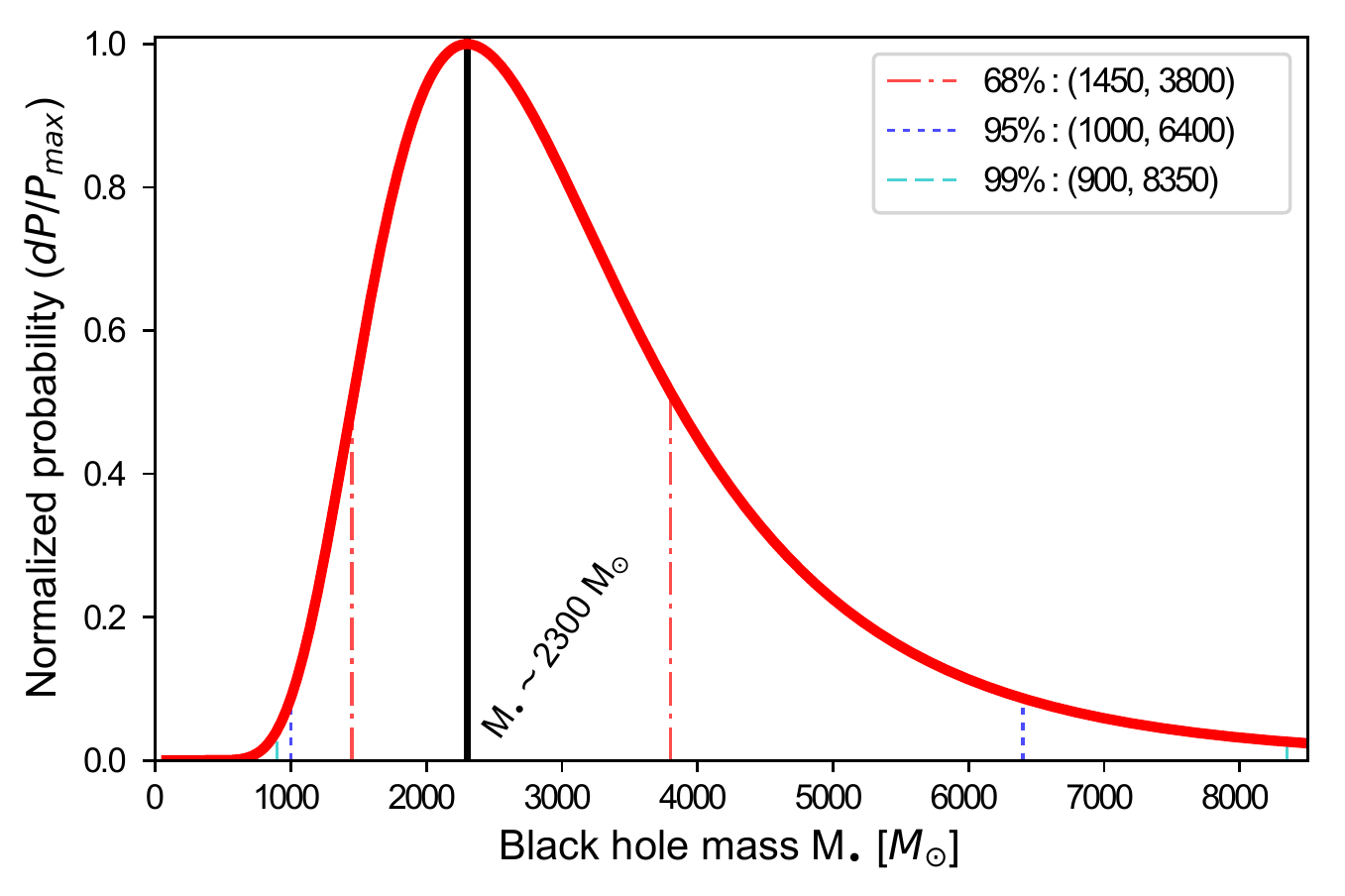} \label{Fig:IMBHmass}} \quad
  \subfigure[]{\includegraphics[scale=0.55]{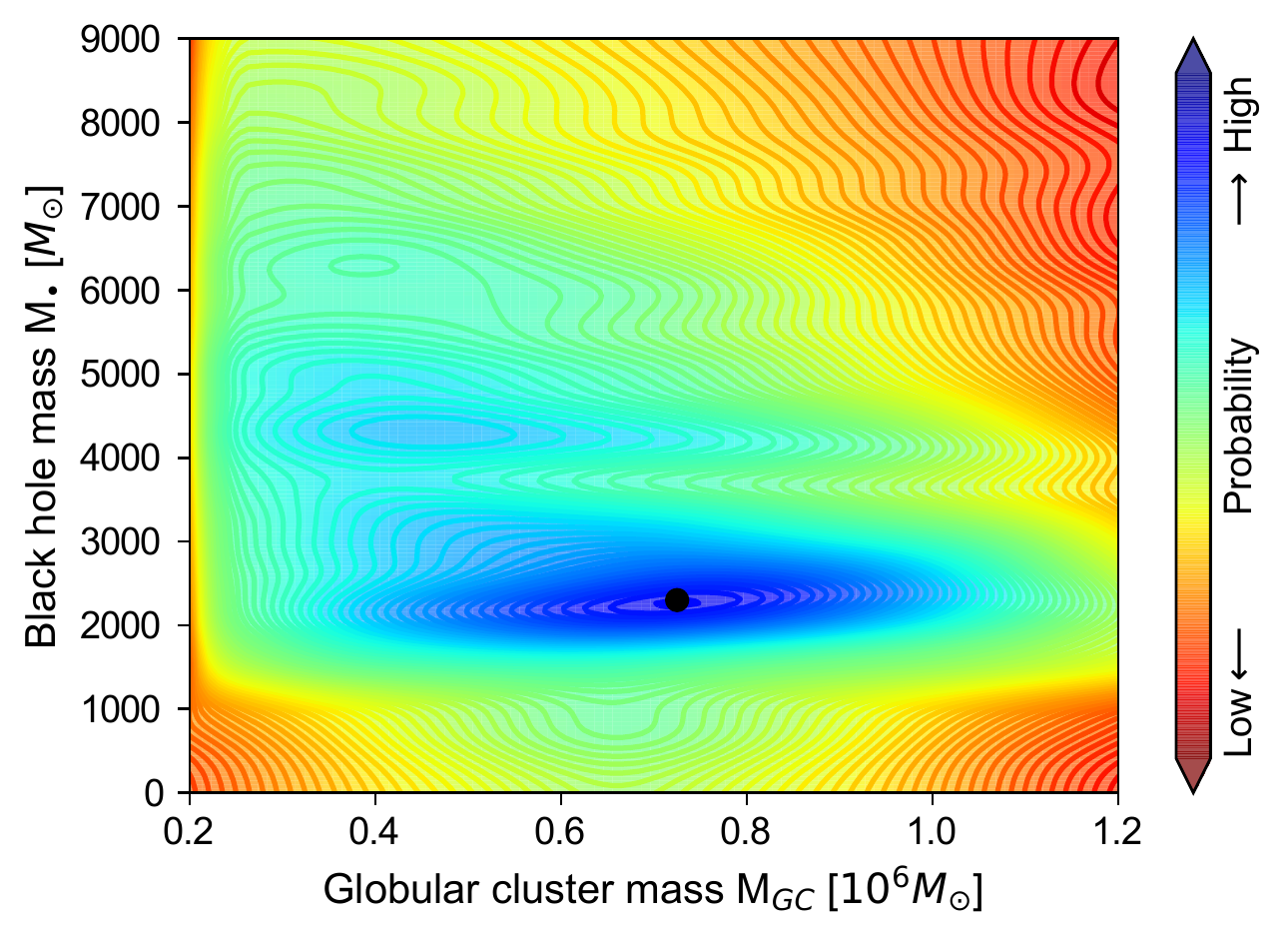}  \label{Fig:IMBHcontour}}
  \caption{{\bf Inferred masses of the black hole and globular cluster.} (a) Pulsars in 47 Tuc imply a central black hole with mass $M_{\bullet}\sim2300M_{\odot}$$_{-850}^{+1500}$ (solid vertical line). The other vertical lines delineate the central 68\%, 95\%, and 99\% estimate limits for the mass of the central black hole. (b) The joint inference for the total cluster mass and the central black hole in 47 Tuc is shown. The black dot shows peak probability at M$_{\bullet}\sim2300 M_{\odot}$ and M$_{GC}\sim0.76\times10^{6} M_{\odot}$}.
\end{figure*}

N-body models with an IMBH produce pulsar accelerations that are overall more consistent with observations compared to models without an IMBH. The relative likelihoods for models in Figure \ref{Fig:models} show that a central black hole is required to produce the observed pulsar accelerations and distributions in 47 Tuc. The inferred black hole mass shows a strong peaks at $M_{\bullet}\sim2300M_{\odot}$$_{-850}^{+1500}$ (Figure \ref{Fig:IMBHmass}) \footnote{Properly scaled models for larger distances (d=4.5 kpc) give slightly more massive IMBHs.}. The total cluster mass is another free parameter in our calculation that serves as a cross check. With this method, we independently measure the total cluster mass of 47 Tuc as $M_{GC}\sim 0.76\times 10^{6}M_{\odot}$ (Figure~\ref{Fig:IMBHcontour}) consistent with kinematic measurements\cite{Watkins:15,Baumgardt:16}.

We employ bootstrapping in order to cross-check whether the black hole detection signature comes from pulsar observations and we can rule out systematic effects. Any {\it statistical distance} approach can be employed to calculate differences for two distributions. The Kullback-Leibler (KL) divergence ($D_{KL}$) is particularly well suited and mathematically robust for cases where the information entropy is quantified as a likelihood ($\mathcal{L}$) (see Section~\ref{sec:stat}). KL divergences are calculated for randomly selected pulsars to quantify relative likelihoods for all models. We trace how the inference for the black hole mass changes in Figure \ref{Fig:IMBHmass}. We find that the inference flattens with decreasing number of randomly selected pulsars (Extended Data Figure 3); this indicates that the black hole detection signal is produced by pulsar data. Such behavior is expected only in systems where there is {\it statistical learning}, i.e. the information driving the inference comes from data, and not from random errors or outliers. For the case of 47 Tuc, the dynamical information extracted from $N<$10 pulsars appears insufficient to conclusively infer an black hole mass while the inference is sufficiently informative for $N>16$ pulsars.

Stars are expected to form a density cusp in power-law form $\rho(r) \propto r^{-7/4}$ around the black hole because of short relaxation times in the centres of globular clusters\cite{Bahcall:76}. However, such a sharp cusp is predicted only for idealized model clusters with single mass. The actual shape of the cusp is difficult to predict because of the unknowns in stellar evolution such as the mass function and binary fraction. While the projected density profile is expected to be relatively sharp, this cannot be used as a direct indication of a similar cusp in luminosity because the mass-to-light ratio ($M/L$) is not constant. More realistic multi-mass cluster models find that giant stars should follow a shallower slope $\sim -0.25$ in a typical cluster in projection\cite{Baumgardt:05}. A segregated neutron star and massive white dwarf population lead to a sharp rise of the dark mass making the luminosity cusp shallower. Moreover, the stochastic motion of an IMBH relative to the cluster centre will enhance the flattening of a centrally rising cusp even further. Therefore, it is not surprising that an IMBH in 47 Tuc would produce an optically unresolved shallow cusp.

An inescapable conclusion for having an electromagnetically undetectable black hole in the cluster centre is that the core is devoid of gas within its radius of influence. Winds driven by main sequence stars\cite{Smith:99}, novae\cite{Scott:78}, and M-dwarfs\cite{Coleman:77} contribute to clearing the gas. Additionally, the kinetic energy injected by the pulsars' high energy emission into the intra-cluster environment alone may be sufficient to clear the ionized plasma from the central core\cite{Spergel:91}. A spherically symmetric distribution of pulsars can provide the integrated ram pressure enhancement in the centre to create a gas free cavity. The observed signature in the dispersion measures of pulsars in 47 Tuc\cite{Freire:01a} is likely due to the bound ionized plasma remaining present beyond the radius of influence of the black hole.

It is possible that such black holes sitting in the gas deficient central cavities of primordial globular clusters constitute a subpopulation of progenitor seeds that formed the supermassive black holes in galaxy centres. For this, the black hole must form rapidly during the early evolutionary phases of the cluster and continue to grow exponentially after the cluster sinks to the core of a galaxy. Although massive and compact clusters are preferred environments where IMBHs can be found, mounting evidence suggests that the formation and retention of such black holes are stochastic\cite{Giersz:15}. For cases in which seed black holes are retained, it has been shown that they can grow rapidly through accretion at super-Eddington rates\cite{Soria:14}. Stellar winds of massive stars can provide a gas reservoir with $\dot{M}_{wind} \sim 10^{-6}(M_{GC}/10^{5}M_{\odot}) $ yr$^{-1}$ to partially feed the black hole\cite{Pfahl:02}. The gas present in the core can then be accreted as the IMBH stochastically moves in the core due to the encounters. It is likely that runaway collisions\cite{Portegies-Zwart:04} play at least a partial role in the growth of black holes as well as the mergers of compact binaries\cite{Miller:02}. In fact, binaries with a black hole primaries are expected to form in the early phases of the cluster evolution and may be retained\cite{Morscher:15}. Some of these massive binaries can eventually merge and drift toward the centre feeding the black hole making these sources ideal for tidal disruption and gravitational wave detection. The rapid growth of a black hole in a primordial globular cluster that merges with the galactic bulge as it passes close to the centre of a galaxy may continue exponentially at super Eddington rates\cite{Alexander:14} producing the supermassive black holes in quasars\cite{Richstone:98}.

\begin{addendum}
\item [Acknowledgements] This work was supported in part by the Black Hole Initiative at Harvard University, through a grant from the John Templeton Foundation.
\vspace{-.3cm}
\item [Author Contributions] B. K. initiated the project, led the collaboration, and analyzed the data. B. K. and H. B. wrote the manuscript. H. B. calculated the N-body models. A. L. made contributions to the conceptual definition of the project. All authors contributed to the interpretation of the analysis.
\vspace{-.3cm}

\item[Competing Interests] The authors declare that they have no competing financial interests.
\vspace{-.3cm}

\item[Correspondence]  Correspondence and requests should be addressed to B\"ulent K{\i}z{\i}ltan ~(email: bkiziltan@cfa.harvard.edu).
\end{addendum}

\clearpage

\hspace*{-.5cm}{\bf {\Huge Methods}}\\
\vspace{-.5cm}
\section*{Pulsar Data}
\label{sec:pulsardata}

The high precision timing solutions for pulsars in 47 Tuc have been possible due to a long dedicated observation campaign for several decades with the Parkes radio telescope\cite{Camilo:00, Freire:01}. We use the timing solutions in the Extended Table 1{\cite{Freire:03,Ridolfi:16,Pan:16}. 

Based on the observed number of pulsars, the total predicted number of neutron stars in the cluster may change between 200-1500 due to the integrated uncertainties in the luminosity distribution, beaming, flux densities, spectral indices of pulsars; and the initial mass function, binary fraction, encounter rate, scintillation properties of the cluster. Our N-body simulations with a $\sim10\%$ retention fraction predict $\sim$1000 neutron stars in 47 Tuc.

\section*{N-Body Simulations}
\label{sec:nbody}

A grid of several hundred $N$-body simulations of star clusters, varying the initial density profile of the cluster, it's initial half-mass radius and the mass ratio of the black hole to the total cluster mass ($M_{\bullet}/M_{GC}$) were used for our study\cite{Baumgardt:16}. All simulations were produced using the GPU-enabled version of the collisional $N$-body code NBODY6 \cite{Aarseth:99,Nitadori:12} which includes two-body relaxation, tidal fields, fitting formula for the modeling of stellar and binary-stellar evolution. The simulations  were run up to T=11.75 Gyr, the age of 47 Tuc\cite{VandenBerg:13}, and then scaled to have the same half-mass radius as 47 Tuc\cite{Baumgardt:03,McNamara:12,Jalali:12}.The best-fitting models to the surface density and velocity dispersion profiles of 47 Tuc were determined with a $\chi^{2}$ test.

All simulations were run with 100,000 particles except the simulations with 0.5\% IMBHs which used 200,000 particles\cite{Baumgardt:16}. The initial conditions are King (1962) models with initial concentrations of $c$=0.2, 0.5, 1.0, 1.5, 2.0 and 2.5. These values cover the observed concentrations of galactic globular clusters\cite{Harris:96, Harris:10}. Initial relaxation times of the N-body models range from 300 Myr to 12 Gyr with a half-mass radius $R_{H}$=2 pc to $R_{H}$=35 pc. It is unlikely that 47 Tuc has started with a half-mass relaxation time outside this range since this would require that the cluster started with either a mass or half-mass radius very different from that of observed clusters. Also we find that the best-fitting N-body models are well inside the trial grid away from the edges in the parameter space. All models were isolated. IMBHs influence mainly the core of a star cluster and the dynamics in the core are hardly influenced by tidal effects. Therefore, the use of isolated models is justified. 

The black hole was treated as a massive star in the simulations. So the IMBH was not fixed in these simulations and was allowed to wander stochastically around in the core. All models became mass segregated, the amount of mass segregation depending on the initial half-mass relaxation time and initial concentration of the model. During the simulations, stellar mass black holes form from stellar evolution, concentrate near the IMBH due to mass segregation, and then kick each other out due to close encounters. Stellar mass black holes have kicked each other out by T=11.75 Gyr in all simulations, except those with the largest relaxation times where only a handful remain.

Models with black hole masses of 1\% of the cluster mass deviate significantly from the observed density profile of 47 Tuc. The discrepancy is strongest for radii around 10$''$ where the observed density profile has a near constant density core. A near constant density profile at this distance is in good agreement with theoretical predictions\cite{Baumgardt:05}. The 1\% IMBH model also has a strong central rise in the velocity dispersion profile which is not seen in the observations. We expect that both problems would become even more pronounced for more massive black holes. Hence the presence of black holes with mass equal to or larger than $\sim$1\% of the cluster mass ($\sim$8000 $M_{\odot}$) can already be excluded. The 0.5\% IMBH model leads to a better fit of the observational data. At the core of the cluster within 5$''$, the surface density of the 0.5\% IMBH model is compatible with the observed density profile. Only at distances beyond r$\sim$10$''$ the surface density of the model cluster is $<$20\% lower than the observed profile. Such differences could be attributed to the uncertainties in the mass function and binary fractions in 47 Tuc\cite{Baumgardt:04}.

\section*{Primordial Binaries}
\label{sec:primordial}

Primordial binaries are expected to play an important role in the early evolution of the cluster core\cite{Goodman:89}. For 47 Tuc, it has been estimated that primordial binaries may have pushed the core radii up to an additional 20\% during the early stages of the evolution\cite{Giersz:11}. 

In order to comparatively study the possible effects of primordial binaries onto the current mass segregation, we ran additional simulations for three different model clusters. The mass function, neutron star and black hole retention fractions were kept consistent with other runs. These simulations' base code was the GPU enabled NBODY6 with a Hermite integration scheme for variable time steps\cite{Lutzgendorf:13}. This integration scheme applies chain regularization\cite{Hurley:00} in order to have sufficient time resolution to follow the tight orbits of binaries and their close encounters with other stars over the cluster lifetime. 

Our first cluster was evolved with 10\% primordial binaries. We took a snapshot of that simulation at T=11.75 Gyr by which time almost half of all stars  left the cluster and the global binary fraction went down to 7\%. The fraction of binaries within 50\% of the projected half-light radius increased to 15\% at this time, fully consistent with observations of 47 Tuc\cite{Albrow:01}. We compare this with snapshots of the other two simulations, one without primordial binaries and the other with a 0.5\% IMBH model. All snapshots are taken at the same time of their evolution and scaled by their projected half-light radii to compare their stellar distributions.

Extended Data Figure 1 shows clearly that the presence of primordial binaries does not  play a significant role in shaping the current mass segregation profile of massive stars. Once the IMBH forms, it becomes the dominant driver of the cluster core dynamics.

\section*{Information Theory and Statistical Learning}
\label{sec:stat}

The Kullback-Leibler (KL) divergence ($D_{KL}$) provides a mathematically robust approach for a comprehensive treatment of the acceleration distributions\cite{Kullback:51}. $D_{KL}$ quantifies the entropy between two distributions. Specifically, $D_{KL}$ is proportional to $-\E$[log$(\mathcal{L})$], where $\E$ is the expectation value and $\mathcal{L}$ is the likelihood. This likelihood value has a trivial numeric relation to the measured ($\mathcal{P}$) and simulated ($\mathcal{N}$) pulsar accelerations, which then can be calculated given that the information entropy is

\begin{equation} 
D_{KL}(\mathcal{P}||\mathcal{N})=\sum\limits_{i}\mathcal{P}(i) \mathrm{log}\frac{\mathcal{P}(i)}{\mathcal{N}(i)}\propto -\mathrm{log}(\mathcal{L}).
\label{Eq:KL}
\end{equation}

\section*{The Intrinsic Spin-down Contribution to Measured Pulsar Accelerations}
\label{sec:spindown}

The measured spin-down rates ($\dot{P}_{m}$) of galactic millisecond pulsars can be used as a proxy to disentangle the overall potential contribution of the intrinsic spin-down ($\dot{P}_{i}$) to the measured apparent accelerations. For pulsars in globular clusters $\dot{P}_{m} \neq \dot{P}_{i}$ because $\dot{P}_{m}$ is modified by the cluster potential. The acceleration due to the cluster potential is 
\begin{equation} 
a_{GC}=\frac{c}{P}\left(\dot{P}_{m}-\dot{P}_{i}\right).
\label{Eq:accel}
\end{equation}

The variation in the pulsar magnetospheric geometry, formation channels and internal structure (i.e., moment of inertia), which determine $\dot{P}_{i}$, are expected to be similar for pulsars in clusters and the galactic disk. Therefore, the $\dot{P}_{i}$ distribution of pulsars in clusters should be comparable to the $\dot{P}_{m}$ distribution of galactic pulsars. Hence, we use the two-dimensional observed period-spin down ($P-\dot{P}_{m}$) distribution of galactic pulsars to infer the possible range for the intrinsic spin-down contribution to pulsars accelerations in 47 Tuc\cite{Kiziltan:09}. The level of uncertainty due to the unknown contribution of the intrinsic $\dot{P}$ to the measured pulsar accelerations are represented with different shades in Extended Data Figure 2. 

Because the  intrinsic $\dot{P}$ of pulsars are not known, we treat the observed range of $P$ and $\dot{P}$ of galactic millisecond pulsars as a distribution of  values to offset the intrinsic contribution to the measured acceleration. Instead of assuming a single acceleration term, we use the 68\%, 95\%, and a 99\% probable intervals for the range of acceleration values corresponding to each pulsar in 47 Tuc.

\section*{The Globular Cluster Centre}
\label{sec:center}

We use the kinematic centre ($\alpha$, $\delta$) = ($0^{h}24^{m}05^{s}.67$,$-72^{\circ}04^{'}52^{''}.62$) of 47 Tuc\cite{McLaughlin:06} for our calculations. The uncertainty in the centre is about $\pm 0.''25$ in each coordinate\cite{Guhathakurta:92,Calzetti:93}. In order to study the potential contribution of an asymmetry to our calculations, we randomly shift the cluster centre within the error box and re-calculate the black hole mass with randomly selected pulsars for many bootstrap cycles. We find that the effect of uncertainty in the cluster centre to the inferred black hole mass is less than 5\%.

\section*{Code Availability}
\label{sec:code}

The simulations\cite{Baumgardt:16} were produced with the GPU-enabled version of the collisional $N$-body code NBODY6\cite{Aarseth:99,Nitadori:12} publicly available online at this \href{http://www.ast.cam.ac.uk/~sverre/web/pages/nbody.htm}{URL}. The authors are happy to provide further details for reproducing the simulations. 
\vspace*{1.cm}


\begin{table*}
\normalsize
\centering
\begin{tabular}{l l l l l l l} 
\hline 
Pulsar & $\alpha$: RA(J2000) & $\delta$: Dec(J2000) & Distance from &  Period (P) & dP/dt & Ref. \\ 
  &  [hms] & [dms] &  the cluster center & [ms] & $\times 10^{-20}$ [s/s]  & \\
\hline 
\noalign{\smallskip} 

0024--7204AB & 00:24:08.1657(4) & --72:04:47.616(2) & 	0$'$12.6$''$ &  3.7046395539391(2) & +0.9844(3) & \cite{Pan:16}\\ 
0023--7204C & 00:23:50.35311(9) & --72:04:31.4926(4) &	1$'$13.8$''$ &   5.7567799955132(2) & --4.9850(6) & \cite{Freire:03}\\
 0024--7204D & 00:24:13.87934(7) & --72:04:43.8405(3) & 	0$'$38.9$''$ &  5.3575732848627(2) & --0.3429(7) & \cite{Freire:03}\\ 
0024--7205E & 00:24:11.1036(1) & --72:05:20.1377(4) & 	0$'$37.2$''$ &  3.5363291527603(2) & +9.8510(6) & \cite{Freire:03}\\ 
0024--7204F & 00:24:03.8539(1) & --72:04:42.8065(5) & 	0$'$12.9$''$ &  2.6235793525110(1) & +6.4500(4)  & \cite{Freire:03}\\ 
0024--7204G & 00:24:07.9587(3) & --72:04:39.6911(7) & 	0$'$16.7$''$ & 4.0403791435630(4) & --4.215(2)  & \cite{Freire:03}\\ 
0024--7204H & 00:24:06.7014(3) & --72:04:06.795(1) & 	0$'$46.1$''$ &  3.2103407093484(4) & --0.183(1)  & \cite{Freire:03}\\ 
0024--7204I & 00:24:07.9330(3) & --72:04:39.669(1) & 	0$'$16.6$''$ &  3.4849920616611(5) & --4.587(2) & \cite{Freire:03}\\ 
0023--7203J & 00:23:59.40735(3) & --72:03:58.7914(1)&  1$'$01.1$''$ &  2.10063354535247(3) & --0.97921(9)  & \cite{Freire:03}\\ 
0024--7204L & 00:24:03.771(1) & --72:04:56.913(4) & 	0$'$09.8$''$ &  4.346167999460(1) & --12.206(4)  & \cite{Freire:03}\\ 
0023--7205M & 00:23:54.4877(8) & --72:05:30.741(3) & 	1$'$04.2$''$ &  3.676643217598(1) & --3.844(3)  & \cite{Freire:03}\\ 
0024--7204N & 00:24:09.1865(4) & --72:04:28.880(2) &	0$'$28.7$''$ &   3.0539543462594(4) & --2.1870(9)  & \cite{Freire:03}\\ 
0024--7204O & 00:24:04.6512(1) & --72:04:53.7552(5) &	0$'$04.8$''$ &  2.6433432972417(2) & +3.0354(9)  & \cite{Freire:03}\\ 
0024--7204Q & 00:24:16.4891(4) & --72:04:25.153(2) &	0$'$57.0$''$ &   4.0331811845699(5) & +3.402(2)  & \cite{Freire:03}\\ 
0024--7204S & 00:24:03.9779(4) & --72:04:42.342(1) &	0$'$12.9$''$ &   2.8304059578772(4) & --12.054(2)  & \cite{Freire:03}\\ 
0024--7204T & 00:24:08.548(2) & --72:04:38.926(7) &	0$'$19.1$''$ &   7.588479807363(4) & +29.37(1)  & \cite{Freire:03}\\ 
0024--7203U & 00:24:09.8351(2) & --72:03:59.6760(9) & 	0$'$56.3$''$ &   4.3428266963896(4) & +9.523(1)  & \cite{Freire:03}\\ 
0024--7204W & 00:24:06.058(1) & --72:04:49.088(2) &	0$'$04.0$''$ &  2.3523445319370(3) & --8.6553(1) & \cite{Ridolfi:16}\\ 
0024--7201X & 00:24:22.38565(9) & --72:01:17.4414(7) &	3$'$48.6$''$ &   4.77152291069355(5) & +1.83609(7)& \cite{Ridolfi:16}\\ 

\noalign{\smallskip} 
\hline			
\hline
\vspace{.05cm}
\end{tabular}
\caption{\bf (Extended) Pulsars with timing solutions in 47 Tuc (NGC 104)} Numbers in parentheses correspond to uncertainties in the last digit. The period (P) and spin-down (dP/dt) of millisecond pulsars are very precisely measured. The errors quoted in the parentheses are the 1-$\sigma$ errors of the weighted least squares fit to the timing residuals.
\label{table:47Tuc}
\end{table*}

\setcounter{figure}{0}
\begin{figure*}[!t]
\label{Fig:compare}
  \centering
  \subfigure[]{\includegraphics[scale=0.55]{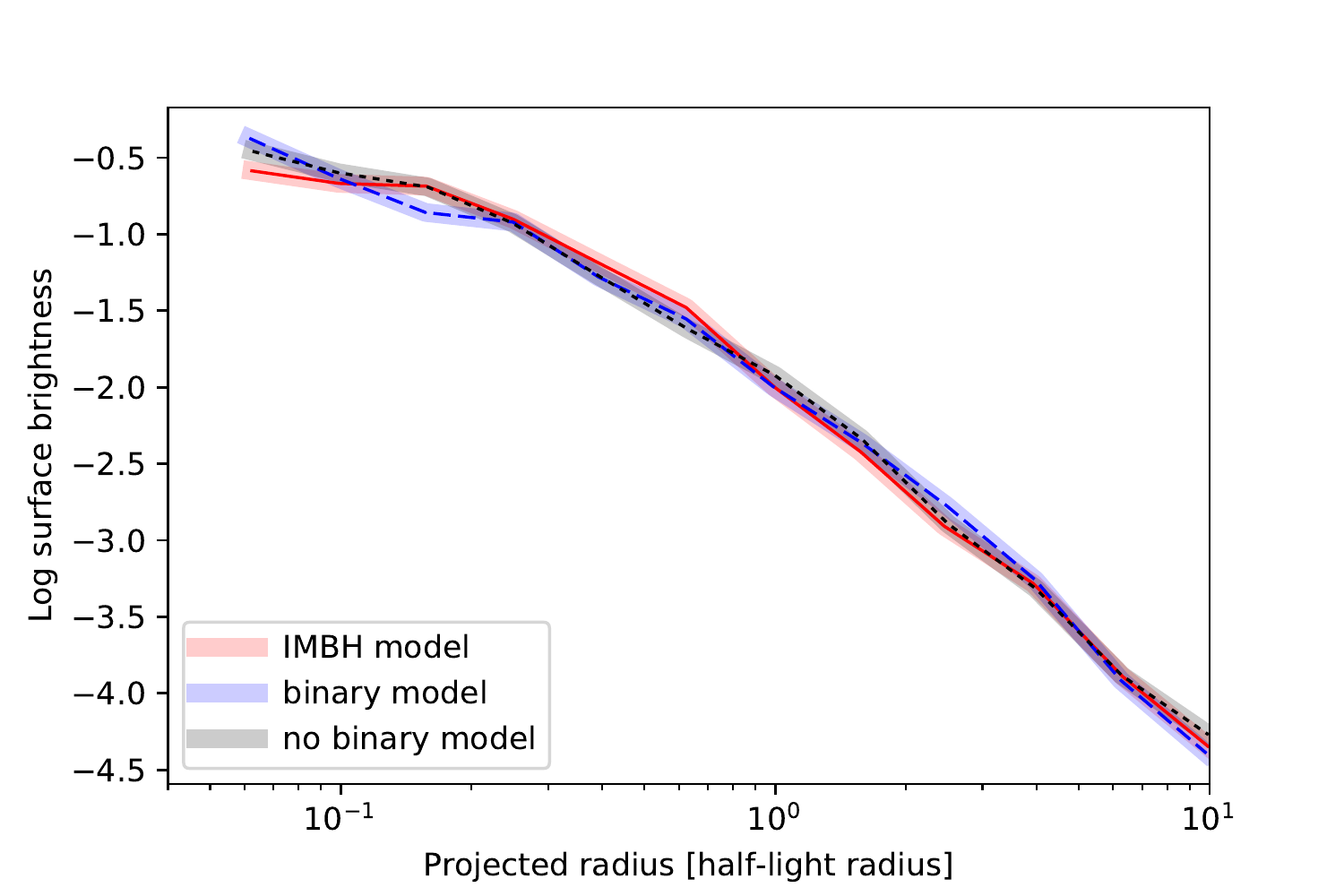}}\label{Fig:surface}\quad
  \subfigure[]{\includegraphics[scale=0.55]{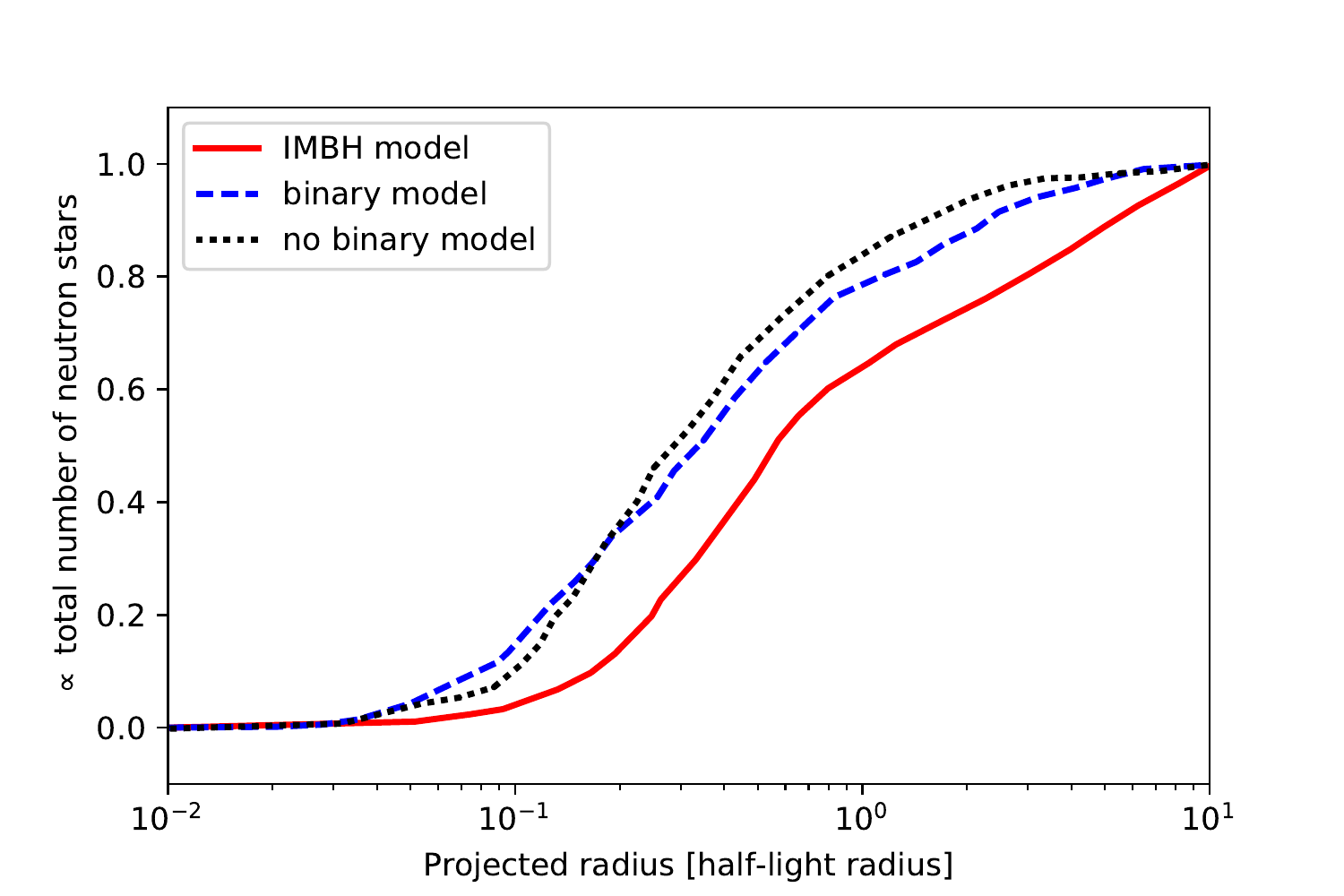}}\label{Fig:primseg}
  \caption{{\bf (Extended) Cluster surface brightnesses and corresponding neutron star spatial distributions for N-body models.} (a) All models have roughly comparable surface brightnesses while (b) the distribution of neutron stars is significantly different for the IMBH model. The neutron star spatial distributions for N-body simulations with and without primordial binaries are similar. Therefore it's unlikely that primordial binaries play a considerable role in shaping the final segregation profile of clusters with an IMBH.}
\end{figure*}

\begin{figure*}
\hspace*{.cm}
\vspace*{-1.5cm}

\subfigure{\includegraphics[trim =  8.mm .0mm .0mm .0mm, clip,width = 1.64in]{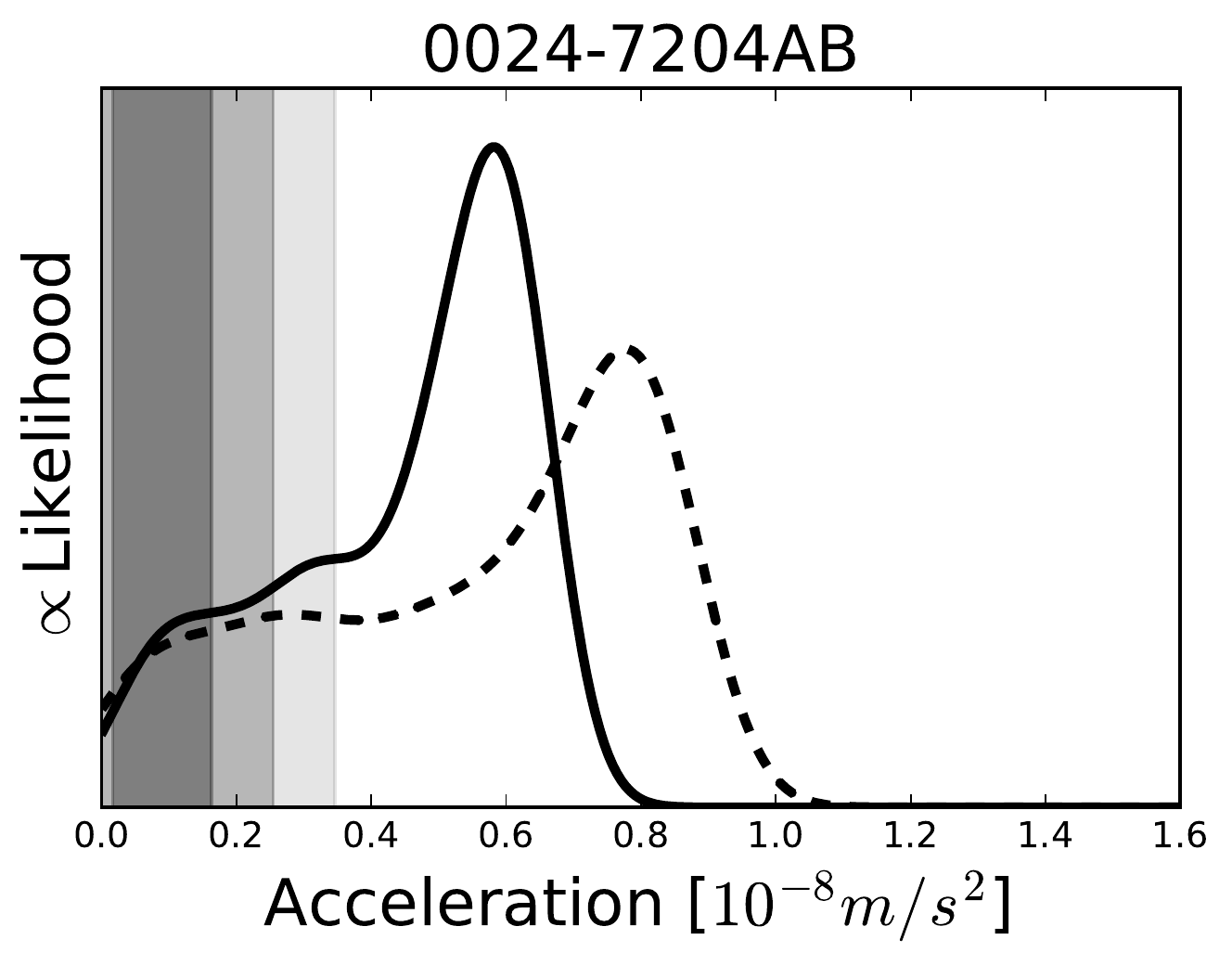}}
\subfigure{\includegraphics[trim =  8.mm .0mm .0mm .0mm, clip,width = 1.6in]{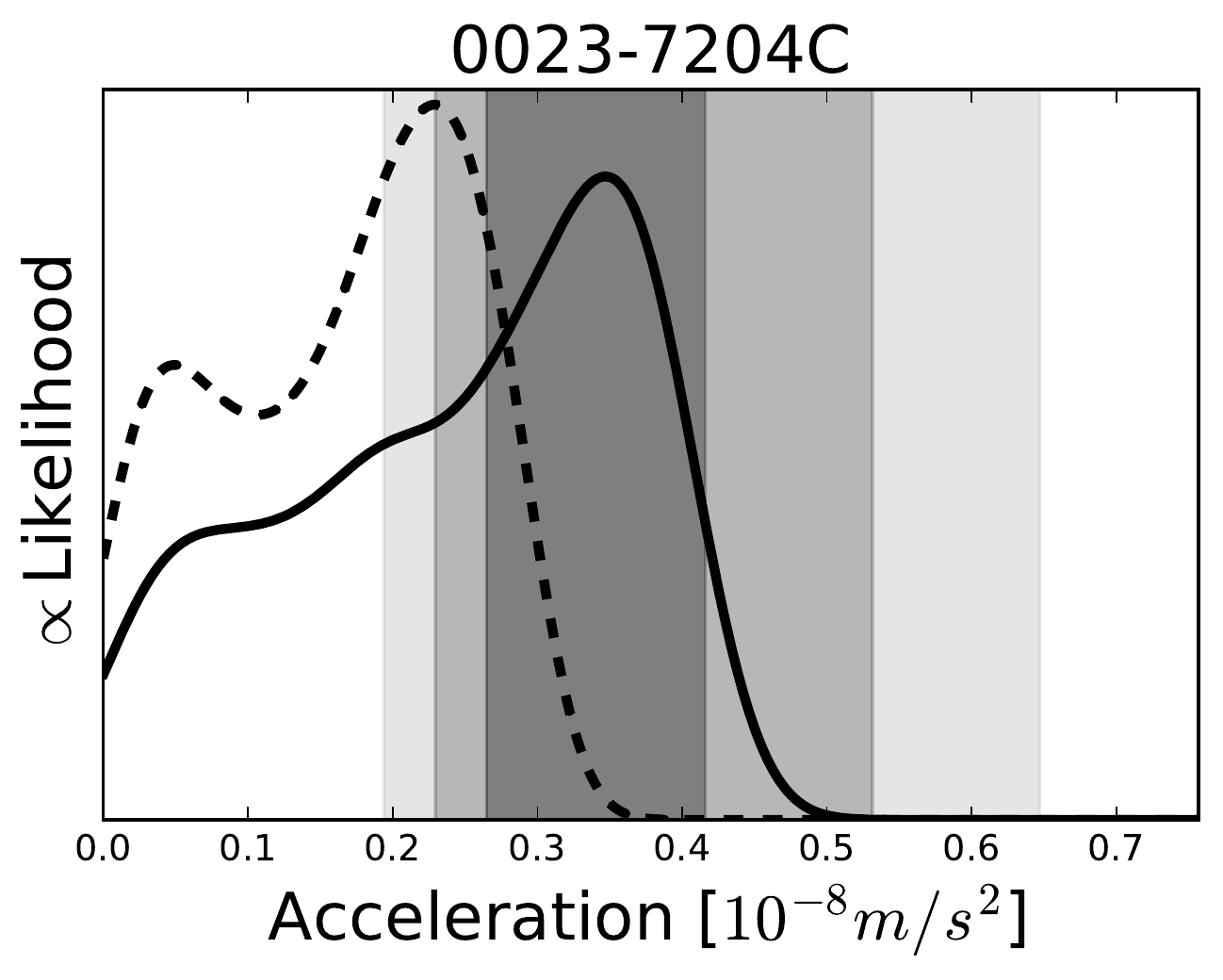}}
\subfigure{\includegraphics[trim =  8.mm .0mm .0mm .0mm, clip,width = 1.6in]{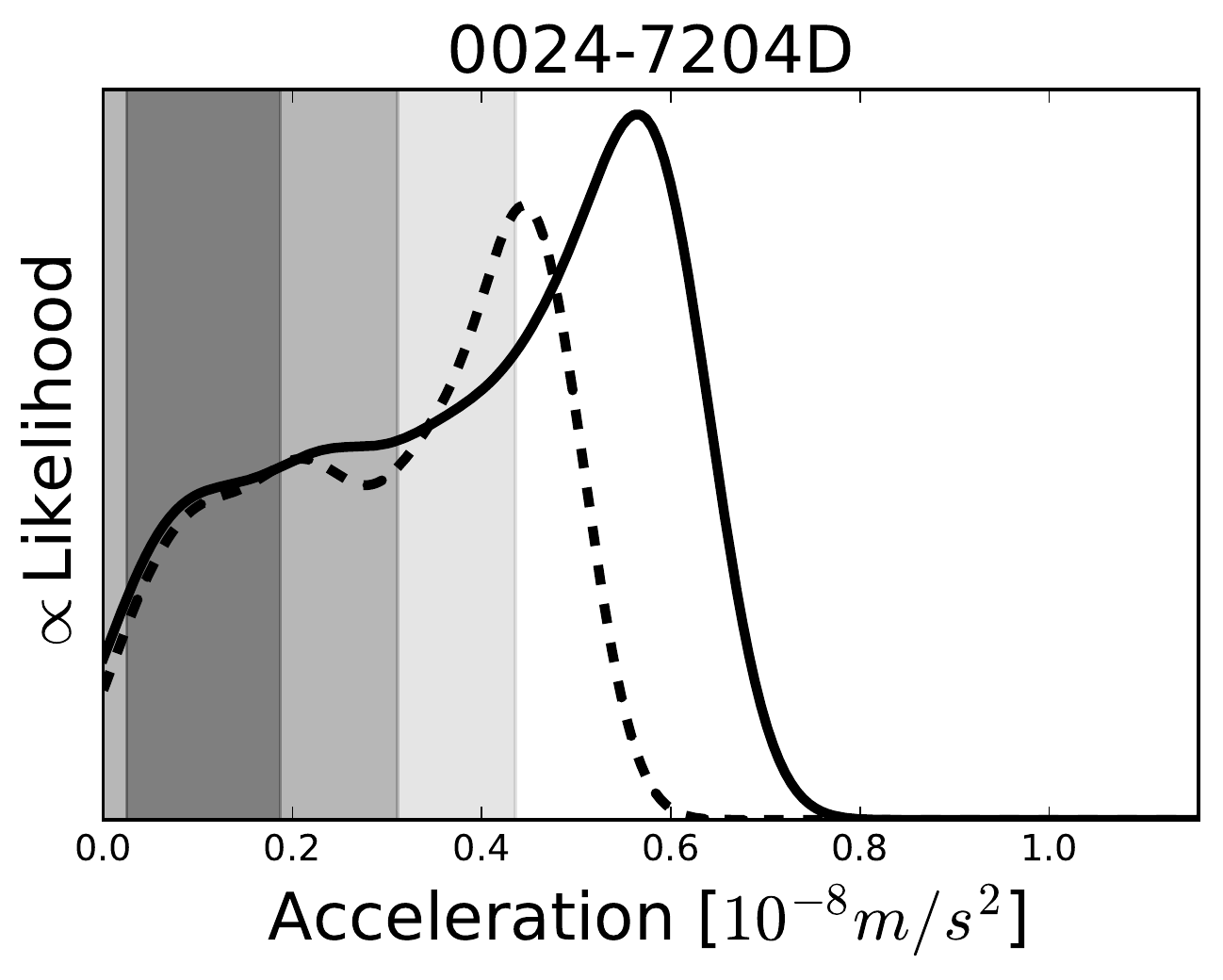}}
\subfigure{\includegraphics[trim =  8.mm .0mm .0mm .0mm, clip,width = 1.6in]{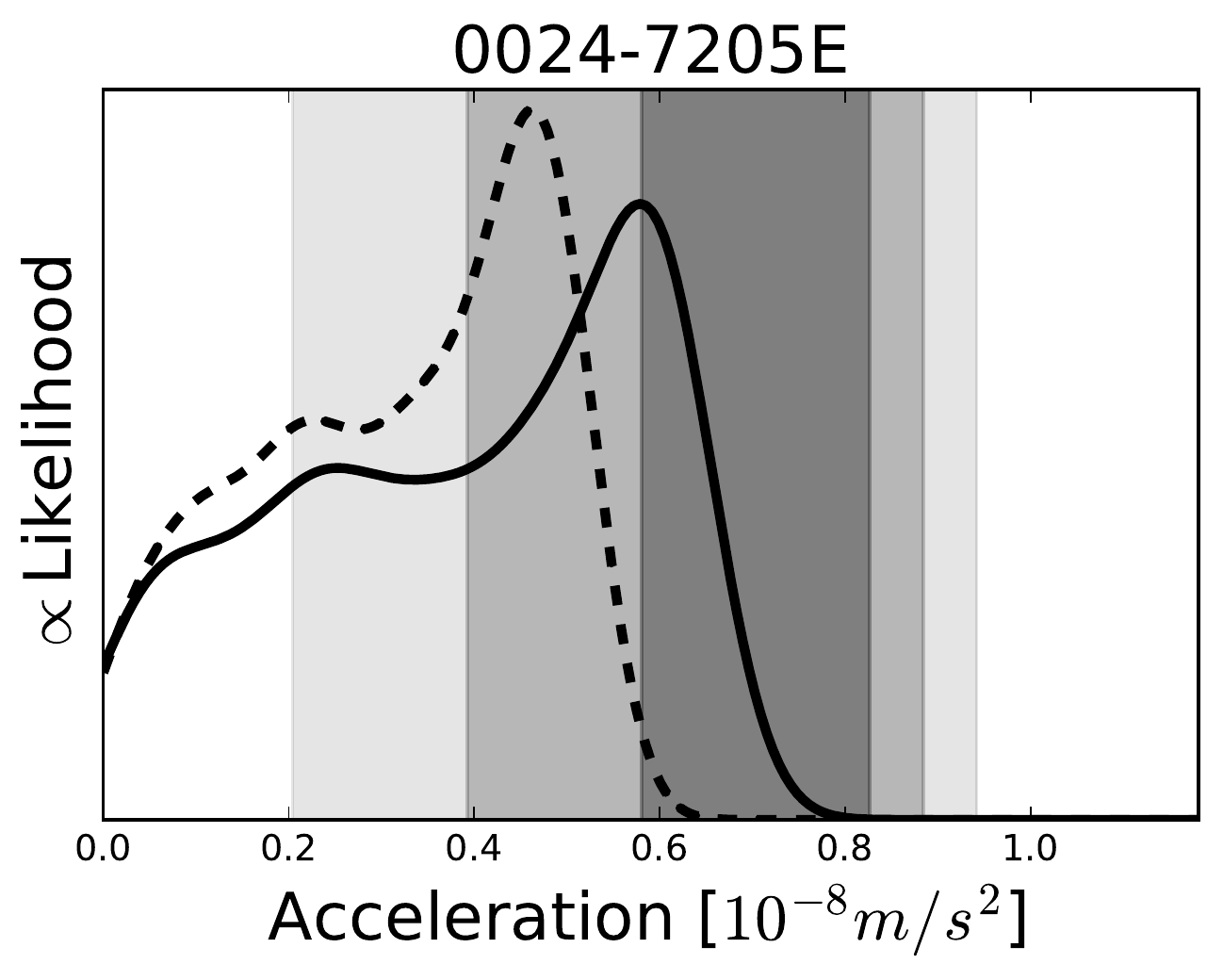}}\\
\vspace*{-.7cm}
\hspace*{.cm}

\subfigure{\includegraphics[trim =  8.mm .0mm .0mm .0mm, clip,width = 1.6in]{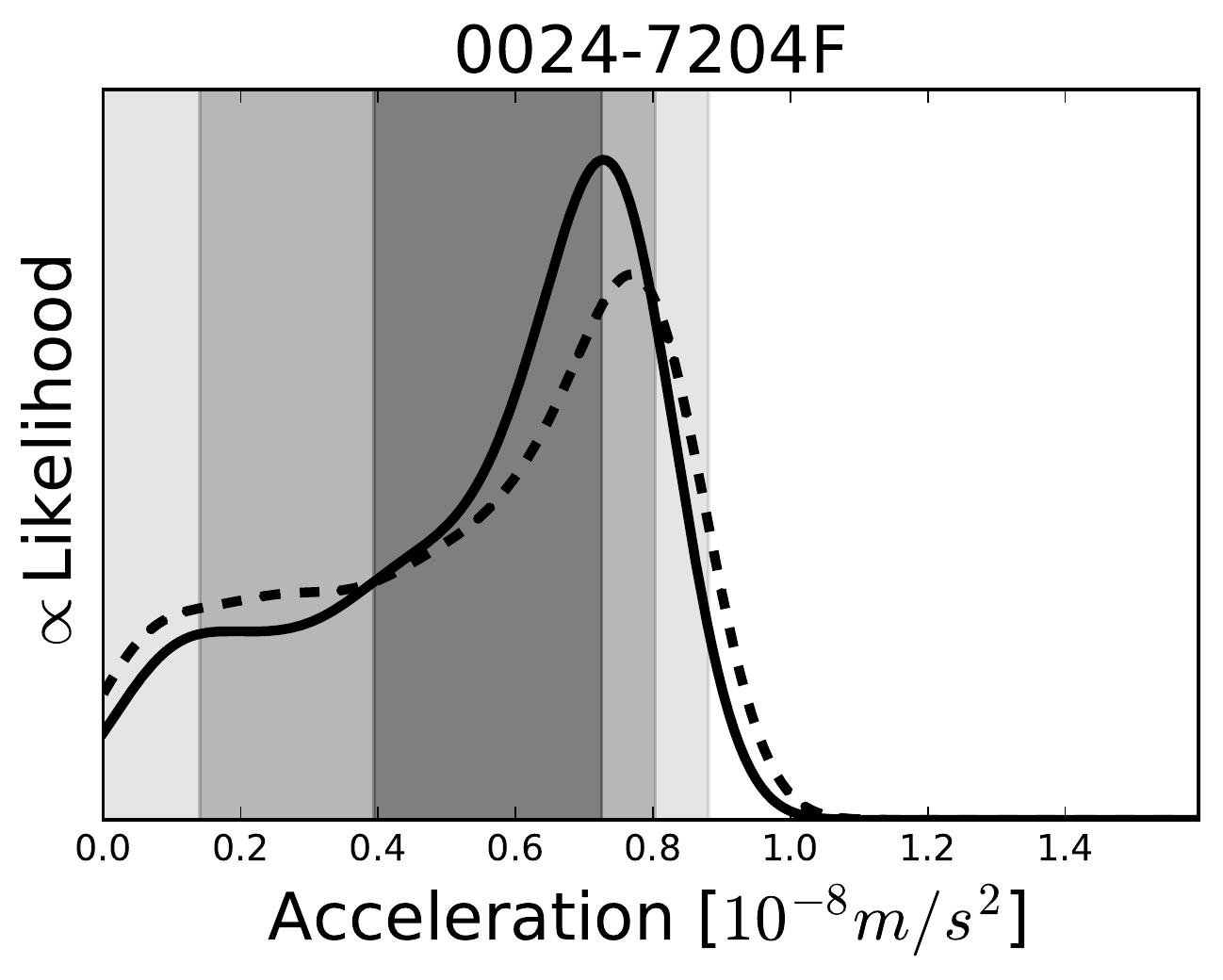}}
\subfigure{\includegraphics[trim =  8.mm .0mm .0mm .0mm, clip,width = 1.6in]{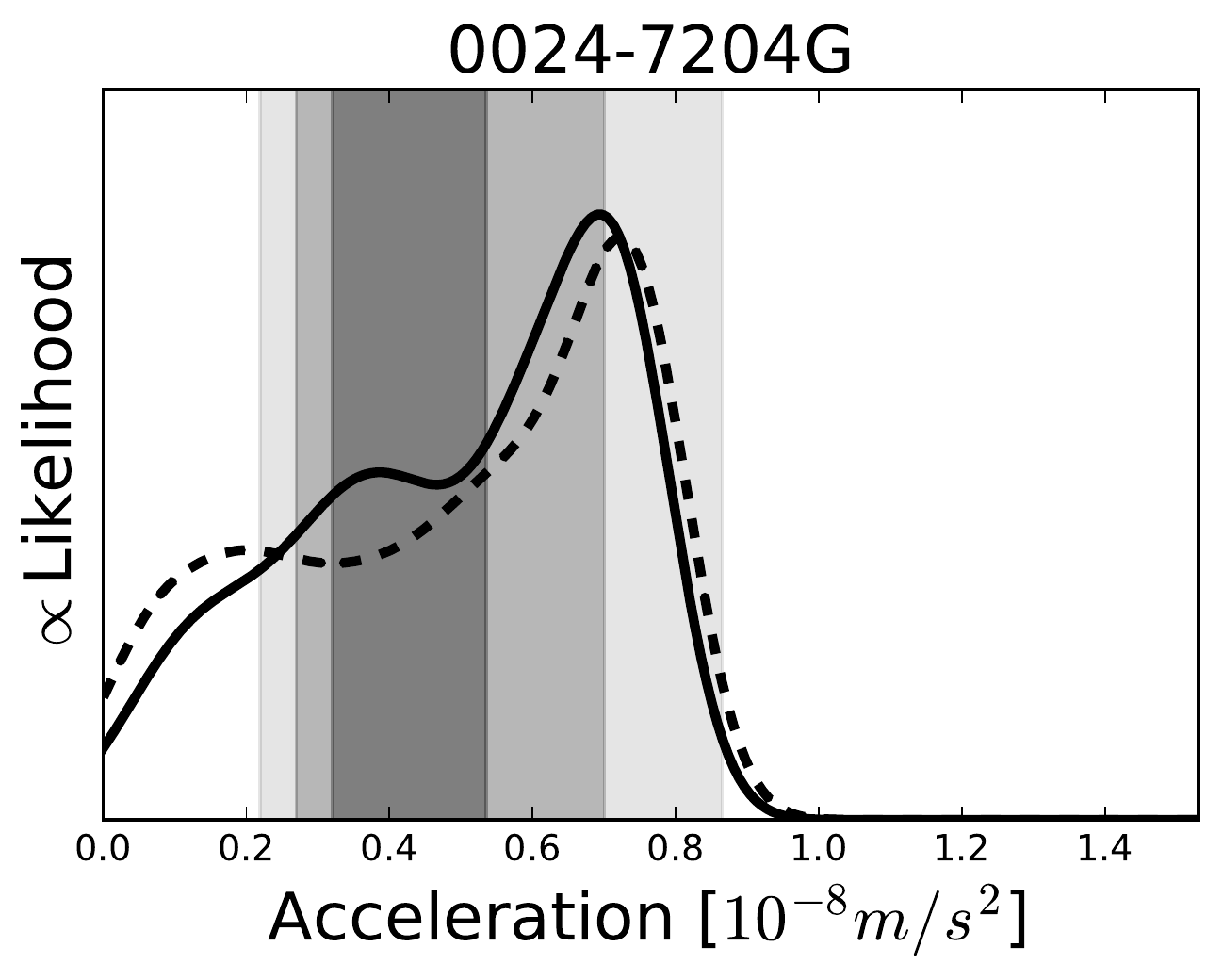}}
\subfigure{\includegraphics[trim =  8.mm .0mm .0mm .0mm, clip,width = 1.6in]{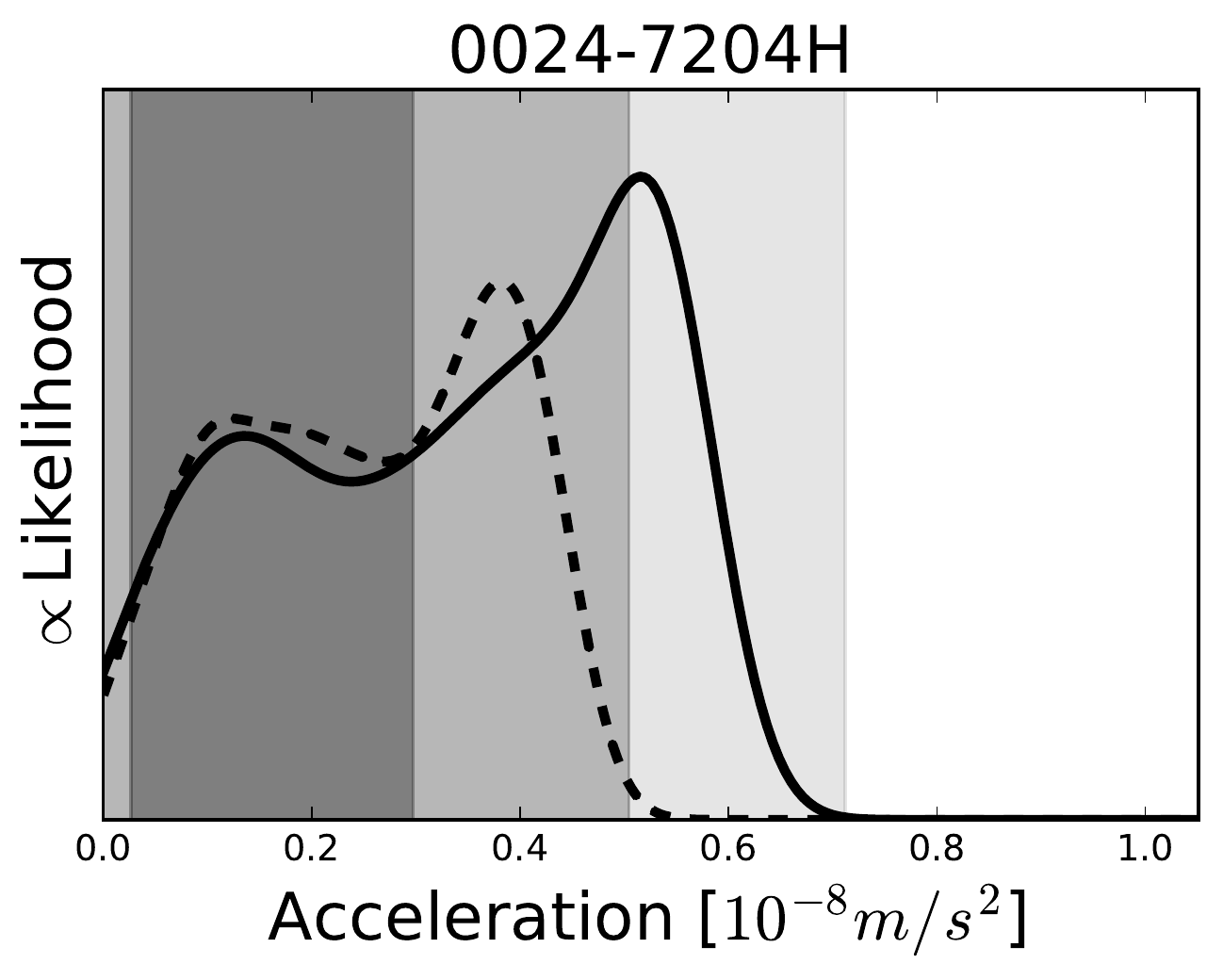}}
\subfigure{\includegraphics[trim =  8.mm .0mm .0mm .0mm, clip,width = 1.6in]{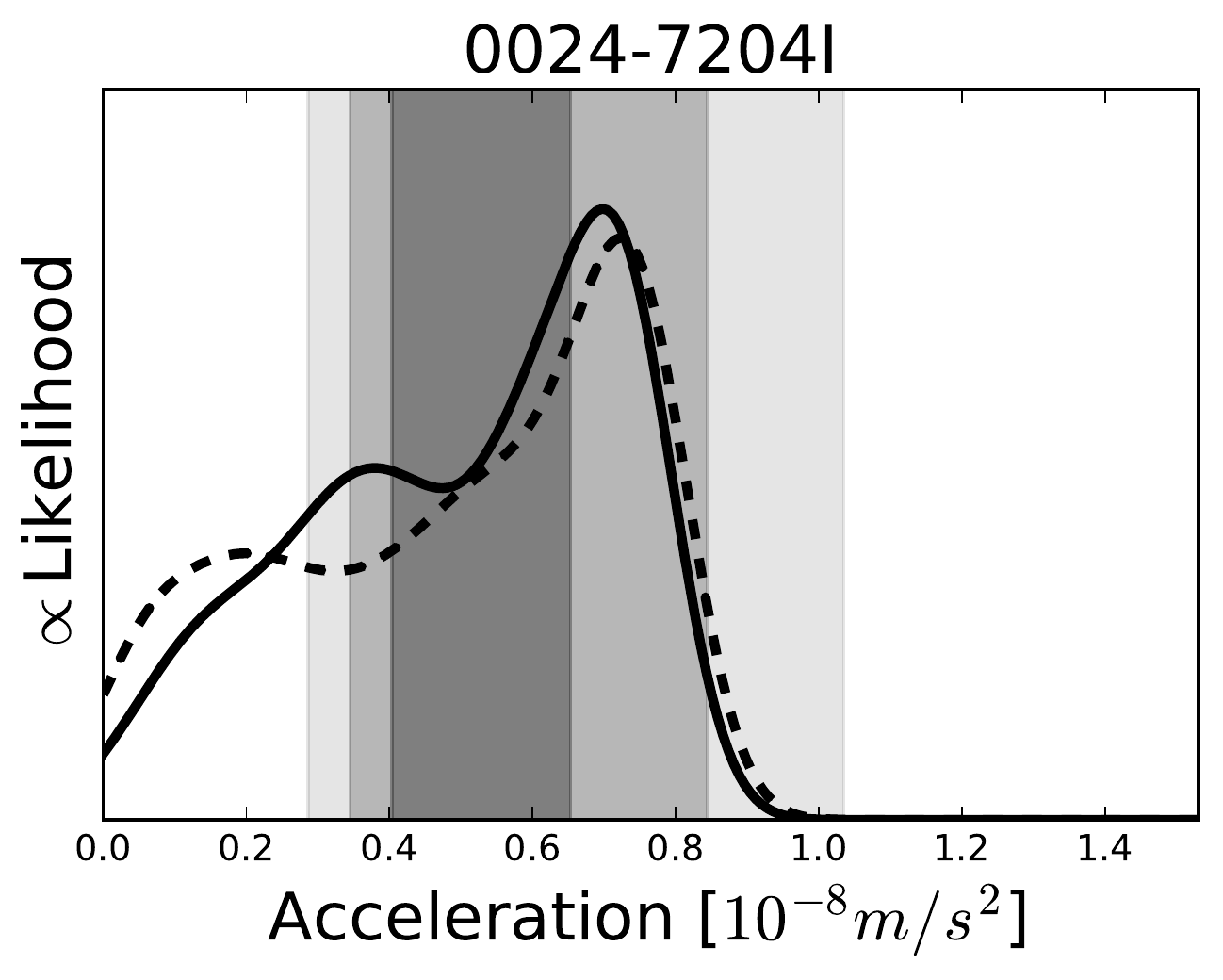}}\\
\vspace*{-.7cm}
\hspace*{.cm}

\subfigure{\includegraphics[trim =  8.mm .0mm .0mm .0mm, clip,width = 1.6in]{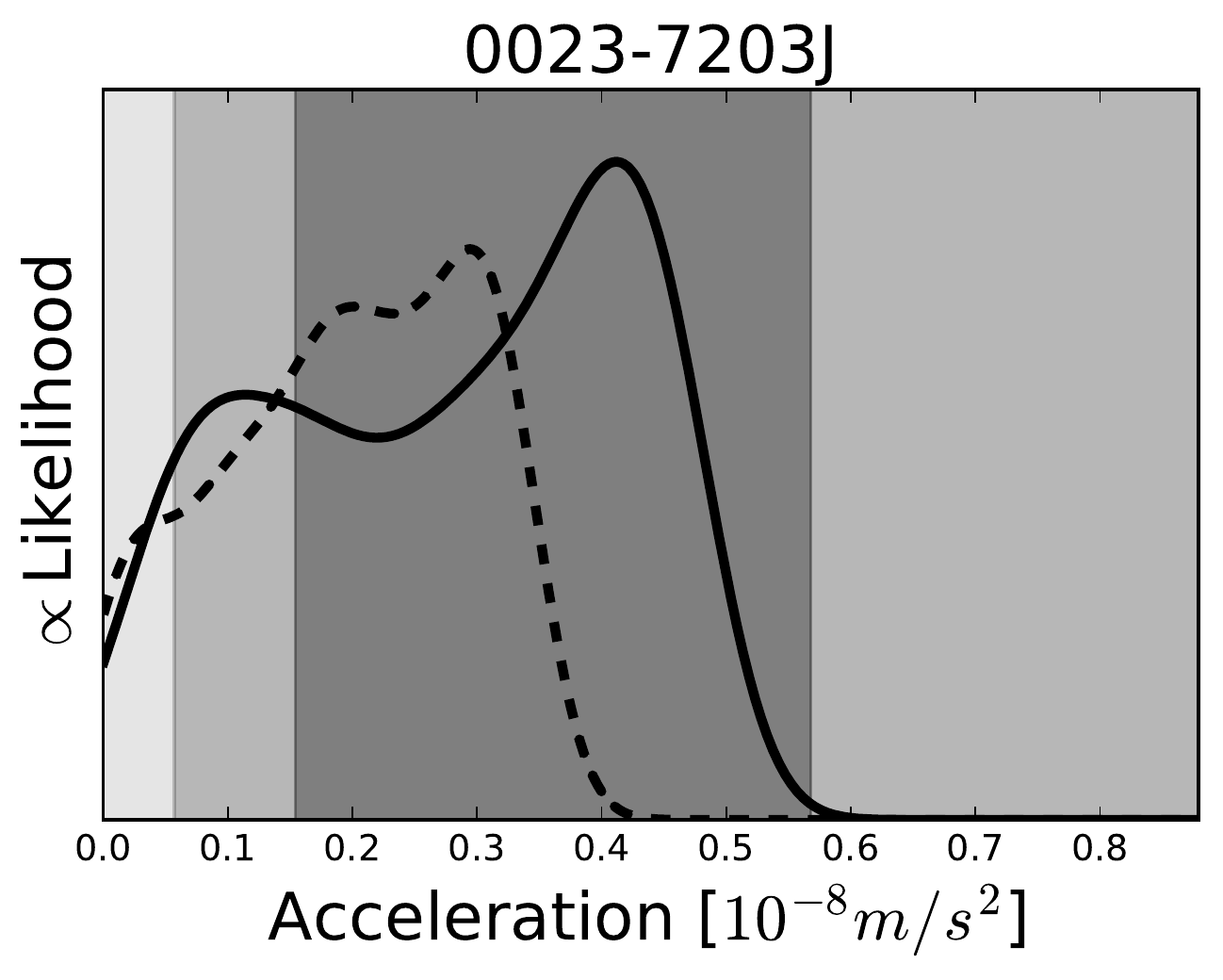}}
\subfigure{\includegraphics[trim =  8.mm .0mm .0mm .0mm, clip,width = 1.6in]{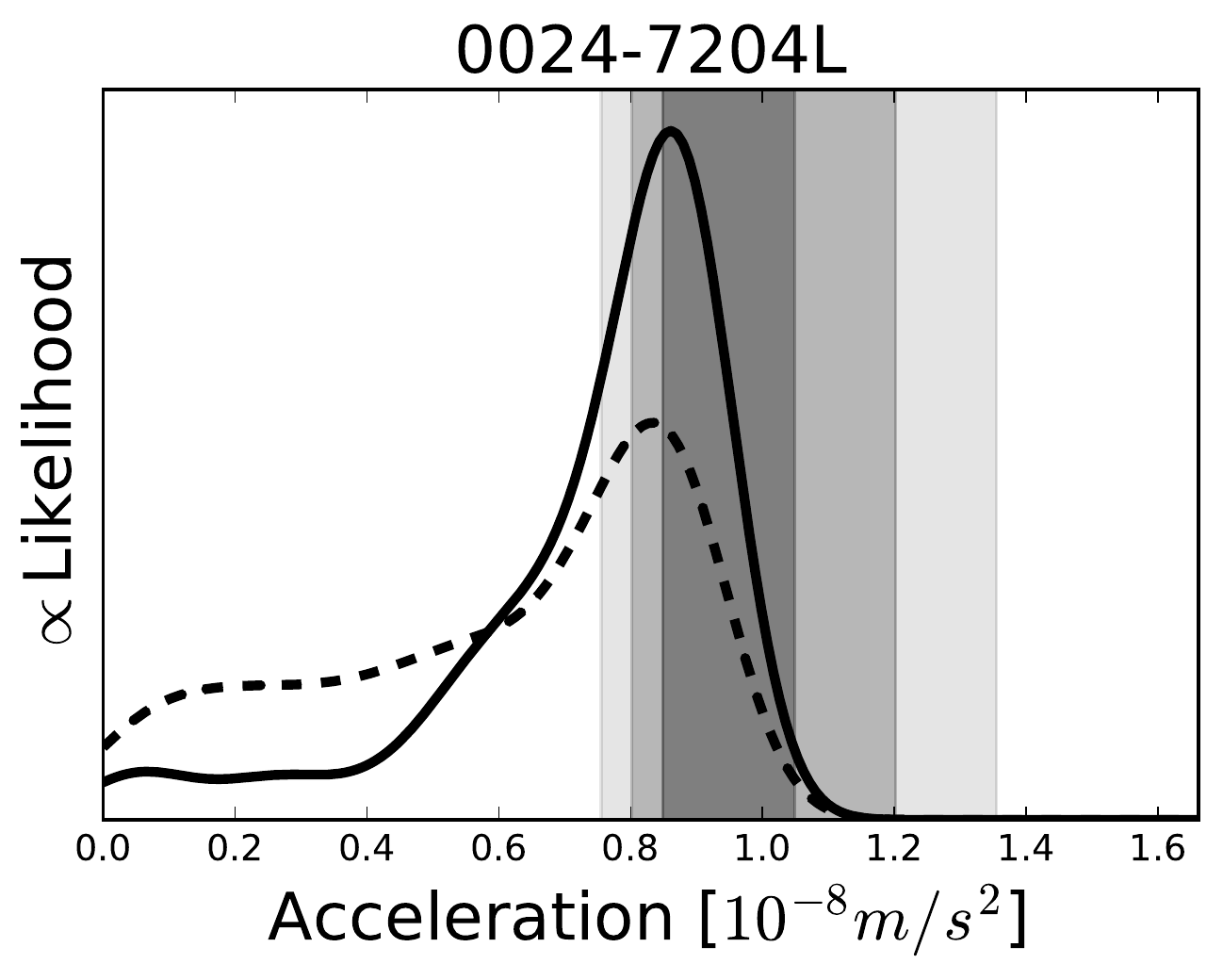}}
\subfigure{\includegraphics[trim =  8.mm .0mm .0mm .0mm, clip,width = 1.6in]{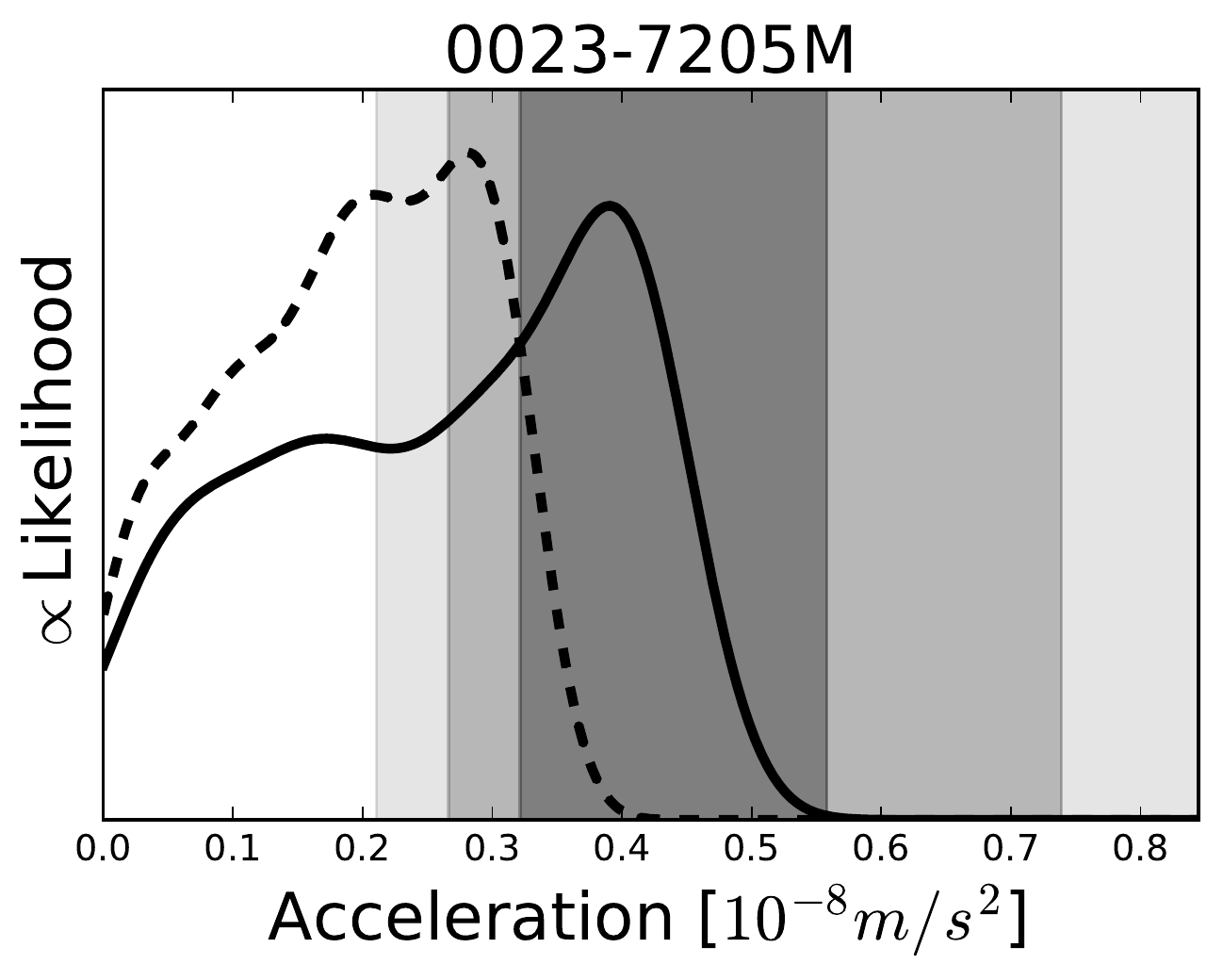}}
\subfigure{\includegraphics[trim =  8.mm .0mm .0mm .0mm, clip,width = 1.6in]{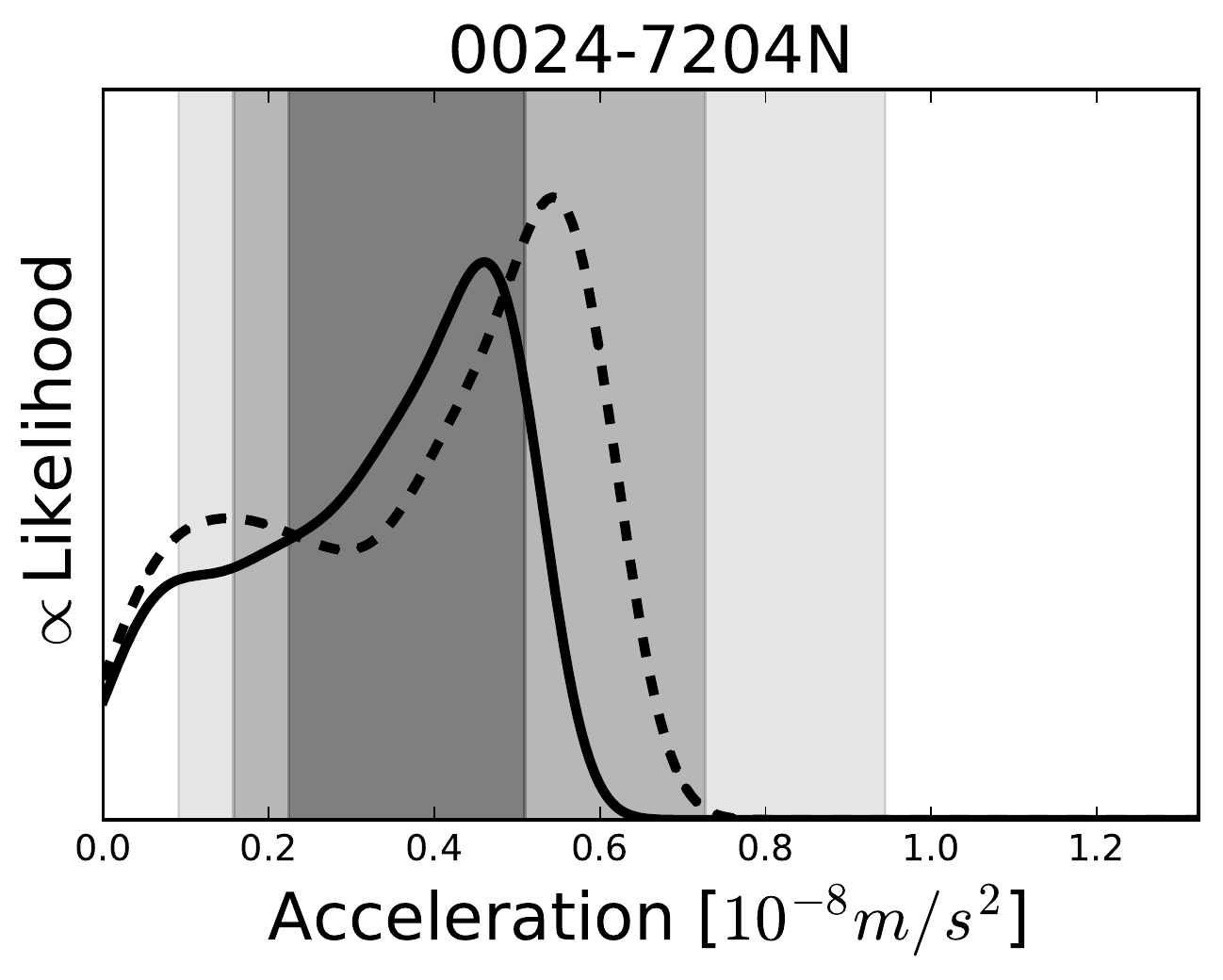}}\\
\vspace*{-.7cm}
\hspace*{.cm}

\subfigure{\includegraphics[trim =  8.mm .0mm .0mm .0mm, clip,width = 1.6in]{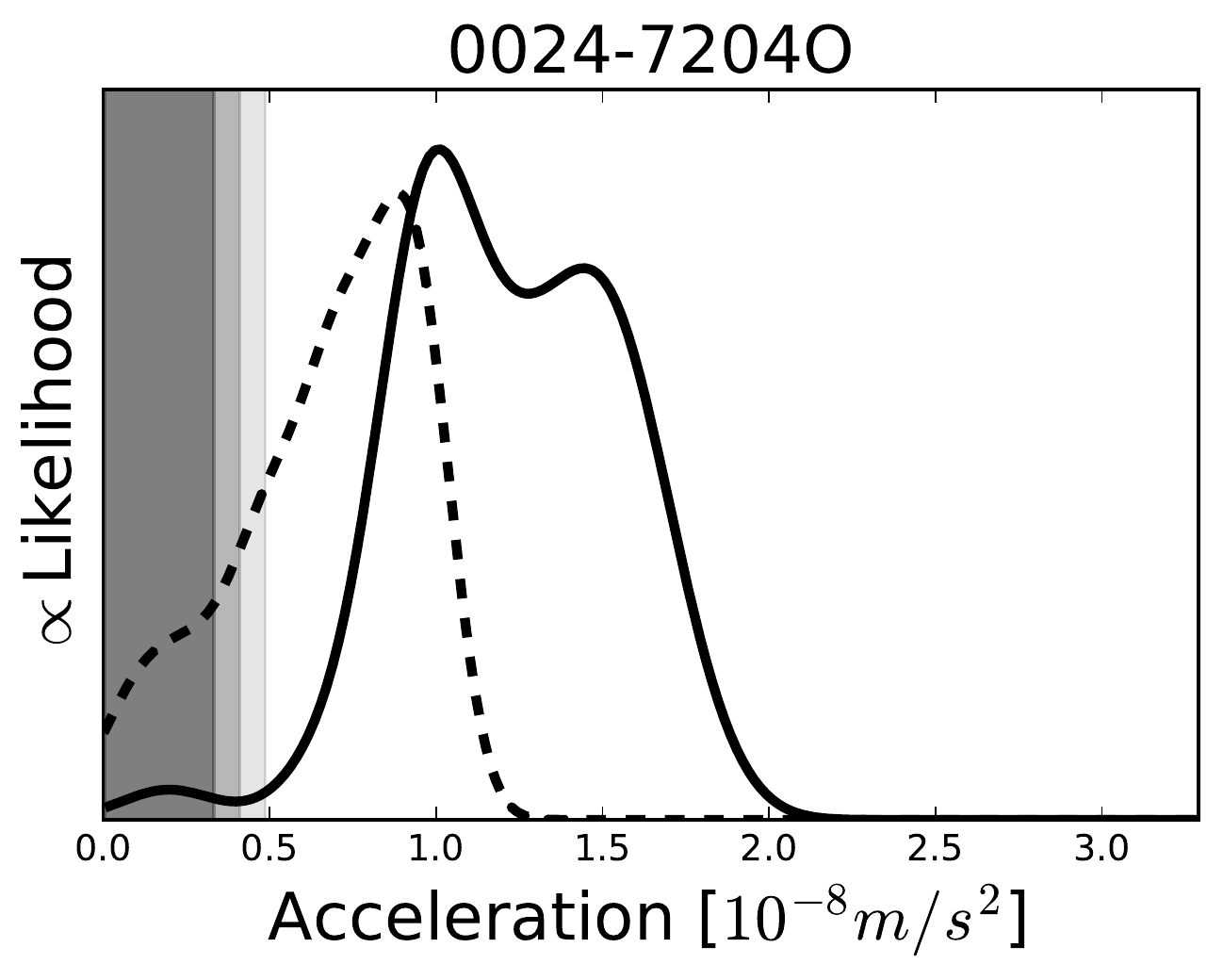}}
\subfigure{\includegraphics[trim =  8.mm .0mm .0mm .0mm, clip,width = 1.6in]{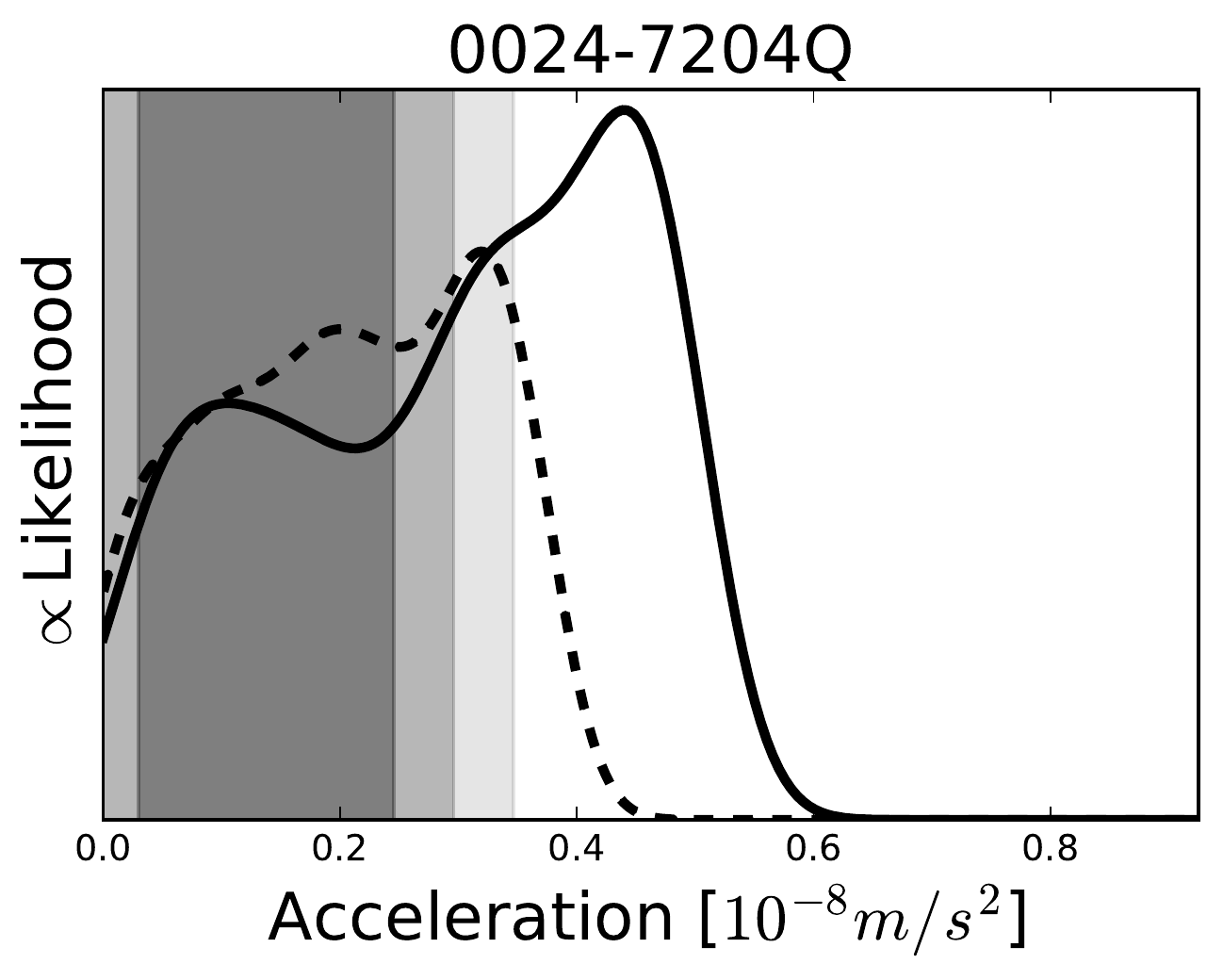}}
\subfigure{\includegraphics[trim =  8.mm .0mm .0mm .0mm, clip,width = 1.6in]{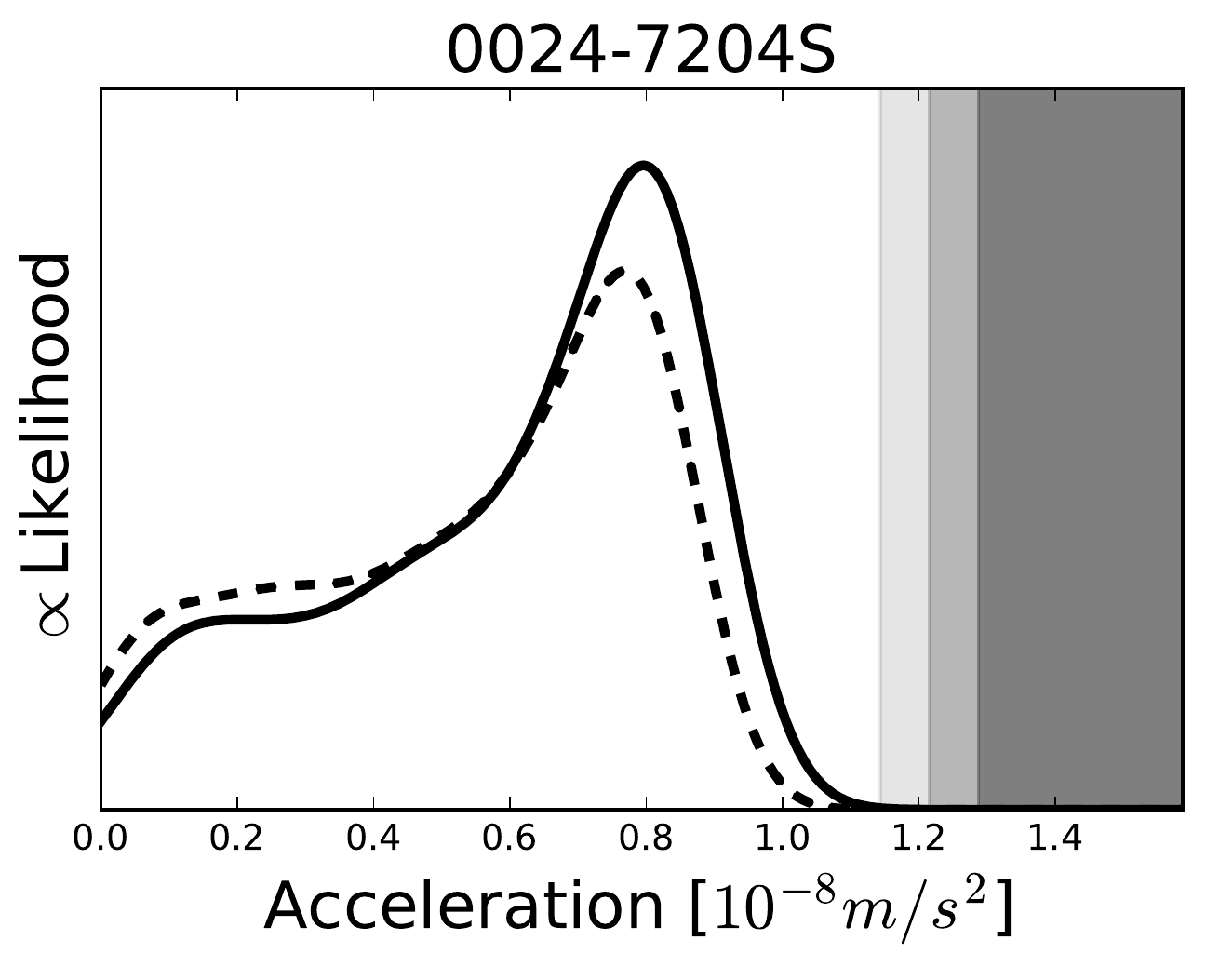}}
\subfigure{\includegraphics[trim =  8.mm .0mm .0mm .0mm, clip,width = 1.62in]{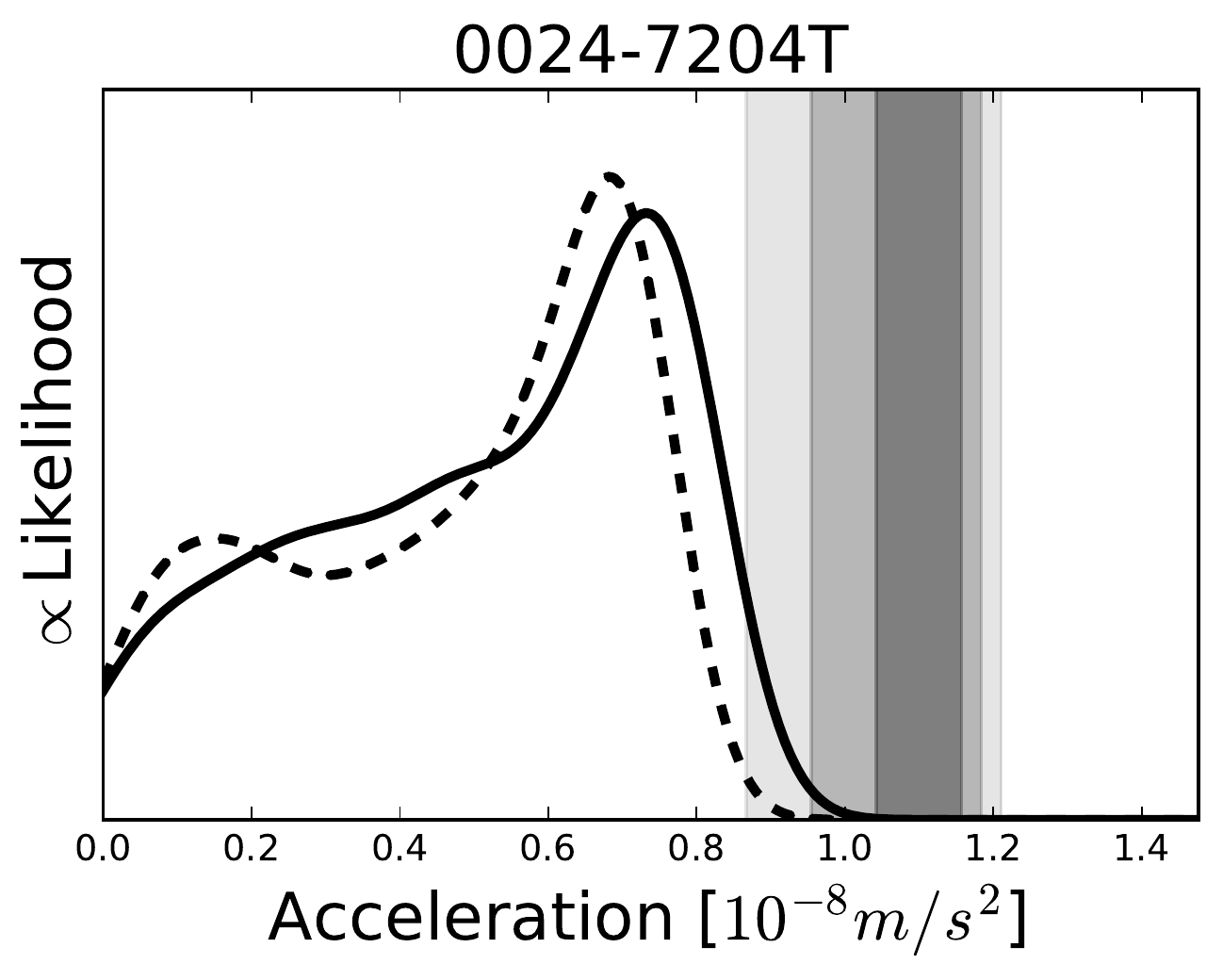}}\\
\vspace*{-.7cm}
\hspace*{.cm}

\subfigure{\includegraphics[trim =  8.mm .0mm .0mm .0mm, clip,width = 1.61in]{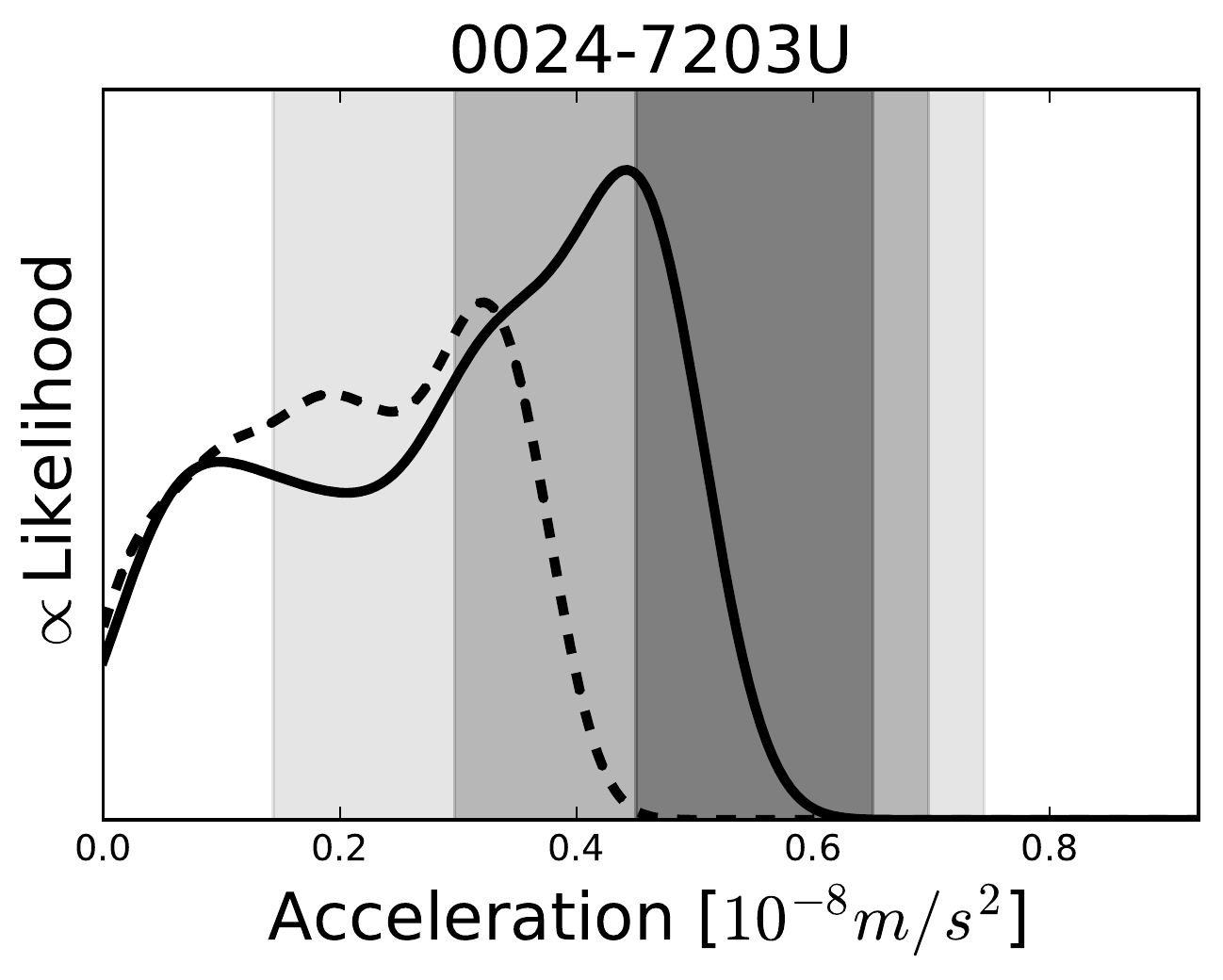}}
\subfigure{\includegraphics[trim =  8.mm .0mm .0mm .0mm, clip,width = 1.6in]{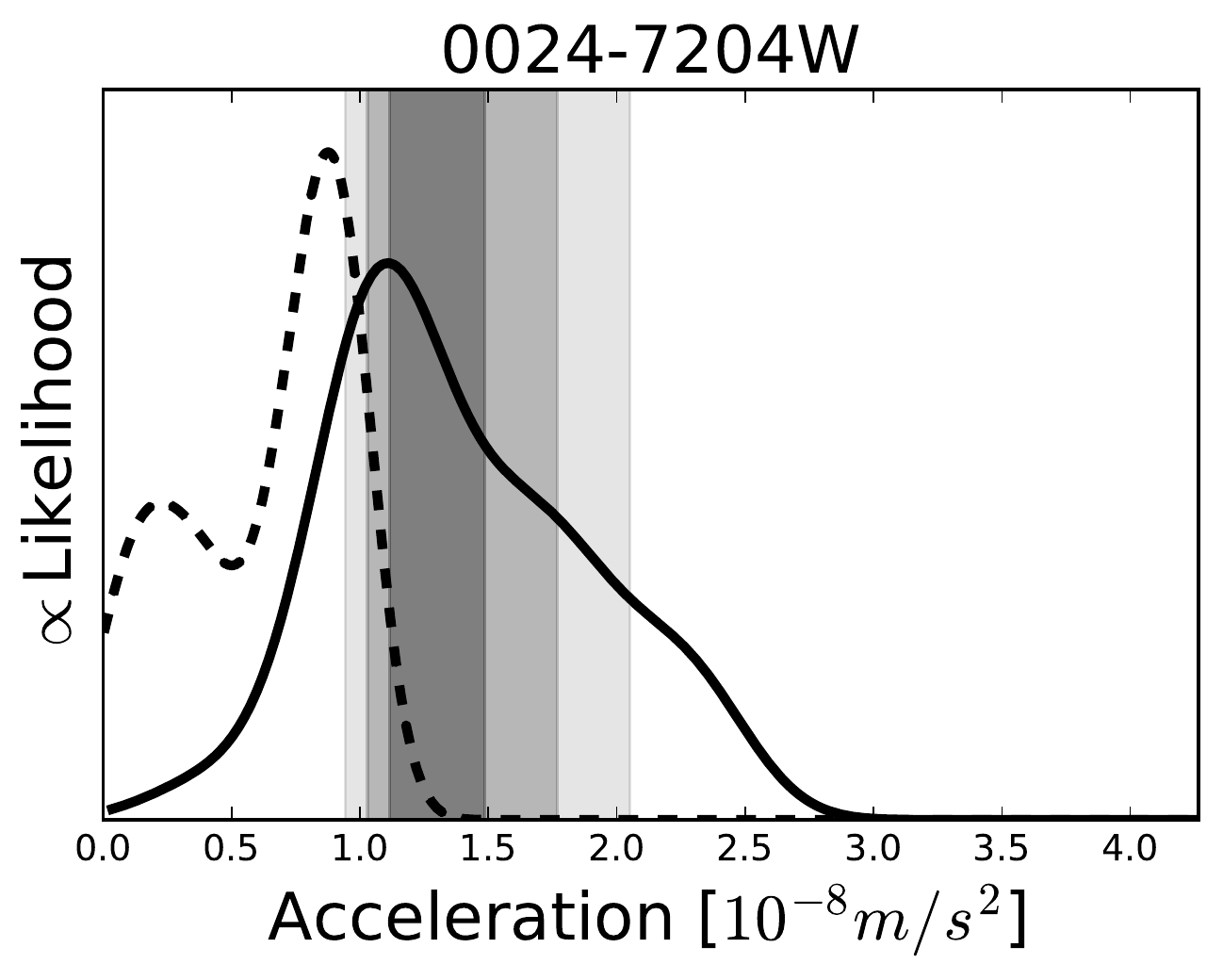}}
\subfigure{\includegraphics[trim =  9.2mm .0mm .0mm .0mm, clip,width = 1.6in]{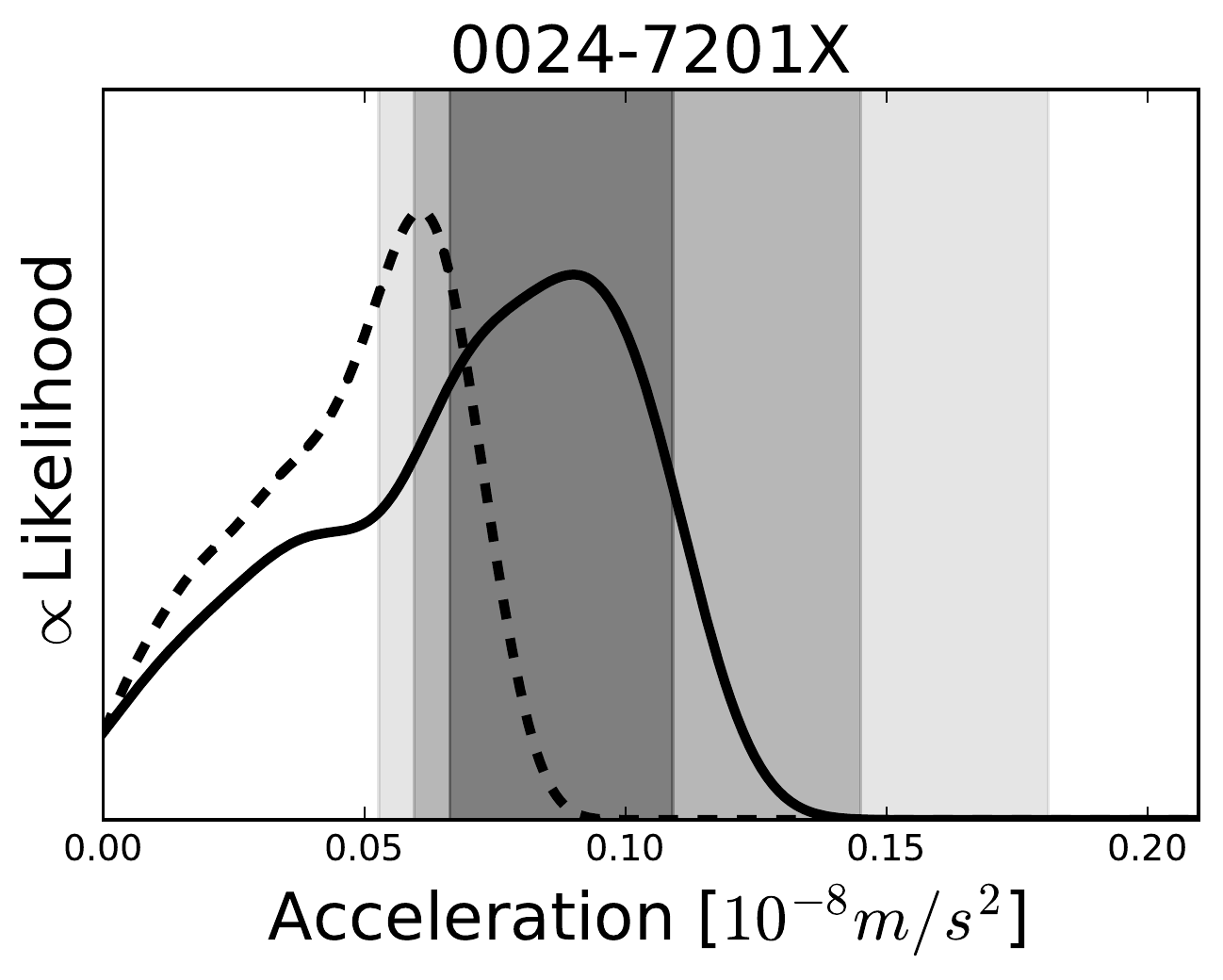}}\\
\vspace*{-.7cm}
\hspace*{.cm}

 \caption{{\bf (Extended) Comparison of the observed and predicted pulsar accelerations.}
The acceleration for each pulsar in 47 Tuc (shaded area) is compared with the integrated acceleration distributions for pulsars with the same line-of-sight distance obtained from $N$-body simulations (models with IMBH-solid line, without IMBH-dashed line). The KL divergence method is used to calculate the integrated information entropy between distributions (Equation \ref{Eq:KL}). The shaded areas show the 68\%, 95\%, and 99\% range of possible accelerations experienced due to the gravitational potential of the cluster. Darker shades represent higher probability. The ambiguity is largely due to the unknown intrinsic spin-down of pulsars.}
\label{fig:acceldist}
\end{figure*}

\begin{figure*}
\centering
\includegraphics[trim =  .0mm .0mm 0.0mm .0mm, clip,width = 5.in]{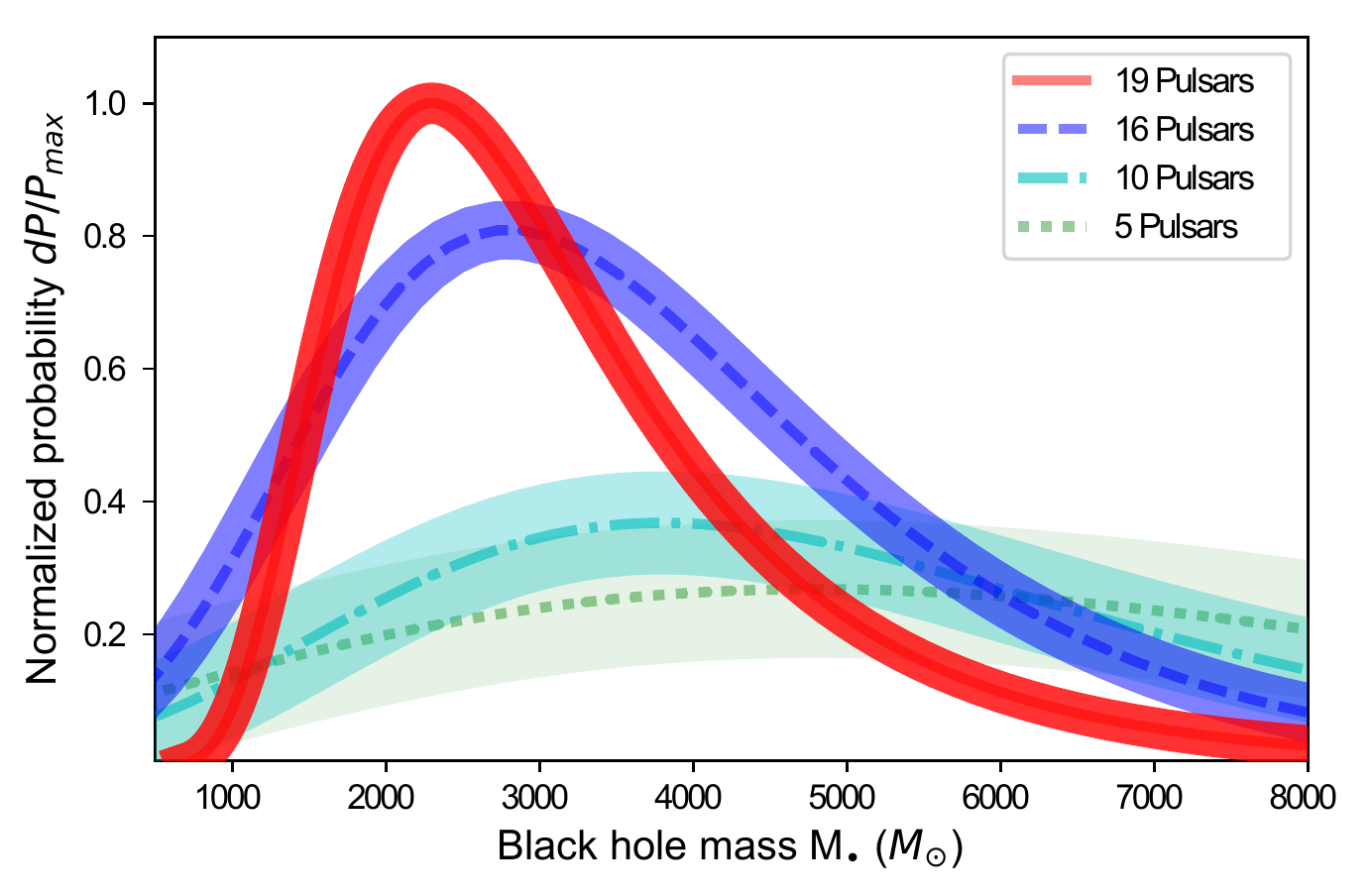}\\
 \caption{{\bf (Extended) Predictive power correlates with number of observed pulsars.} This is the same inference as Figure \ref{Fig:IMBHmass} with different number of randomly selected pulsars. The converging inference with increasing number of pulsars is indicative of {\it statistical learning} and demonstrates that the information comes from observations. The line thickness scales with the level of ambiguity for each inference. }
\label{Fig:learning}
\end{figure*}

{\bf Corrigendum} After the completion of the initial work and submission of this Letter in 2015, some additional data shared with us in 2016 (that we were given to understand will be published shortly) were later incorporated into the analysis owing to an oversight and miscommunication. These data have now been removed from the paper and will be published elsewhere by their originators at a later time. We reran the core analysis on the reduced dataset. Although the main conclusions of the work remain unchanged by these revisions, there are minor changes in the values calculated. Specifically, the central black hole mass changes from 2200$M_{\odot}$ to 2300$M_{\odot}$, the total cluster mass changes from 0.75$\times 10^{6}M_{\odot}$ to 0.76$\times 10^{6}M_{\odot}$, and the number of pulsar timing solutions used in the analysis reduces from 23 to 19. The main text, Figure 4, Extended Data Figs 2 and 3, and Extended Data Table 1 have been updated to reflect these three changes. We apologize for any inconvenience caused by this miscommunication. The original Letter has been corrected online.

\clearpage
\end{document}